\documentclass[twocolumn]{aastex7}
\usepackage{CJK}
\usepackage{amsmath}

\begin{document}
\begin{CJK*}{UTF8}{gbsn}

\title{Galaxy Clusters from the DESI Legacy Imaging Surveys. \uppercase\expandafter{\romannumeral4}. Alignment of Member Galaxy Orientations at $z < 0.5$ } 

\shorttitle{Alignment of Galaxy Orientations}
\shortauthors{Chen et al.}

\author[0009-0006-9628-7318]{Zhaobin Chen}
\affiliation{School of Physics and Technology, Nanjing Normal University, No. 1, Wenyuan Road, Nanjing 210023, People's Republic of China; yuanqirong@njnu.edu.cn}
\affiliation{National Astronomical Observatories, Chinese Academy of Sciences, Beijing 100101, People's Republic of China; zouhu@nao.cas.cn}
\email{chenzbdyx@qq.com}

\author[0000-0003-3196-7938]{Yizhou Gu}
\affiliation{State Key Laboratory of Dark Matter Physics, Tsung-Dao Lee Institute, Shanghai Jiao Tong University, Shanghai 200240, People's Republic of China}
\email{guyizhou@sjtu.edu.cn}

\author[0009-0007-8264-8343]{Shufei Liu}
\affiliation{National Astronomical Observatories, Chinese Academy of Sciences, Beijing 100101, People's Republic of China; zouhu@nao.cas.cn}
\affiliation{University of Chinese Academy of Sciences, Beijing 100039, People's Republic of China}
\email{sfliu@nao.cas.cn}

\author[0000-0002-6684-3997]{Hu Zou}
\affiliation{National Astronomical Observatories, Chinese Academy of Sciences, Beijing 100101, People's Republic of China; zouhu@nao.cas.cn}
\affiliation{University of Chinese Academy of Sciences, Beijing 100039, People's Republic of China}
\email[show]{zouhu@nao.cas.cn}

\author[0000-0002-9244-3938]{Qirong Yuan}
\affiliation{School of Physics and Technology, Nanjing Normal University, No. 1, Wenyuan Road, Nanjing 210023, People's Republic of China; yuanqirong@njnu.edu.cn}
\affiliation{University of Chinese Academy of Sciences, Nanjing 211135, People's Republic of China}
\email[show]{yuanqirong@njnu.edu.cn}

\correspondingauthor{Hu Zou, Qirong Yuan}

\begin{abstract}

Using DESI photometric catalog data, we systematically analyze the spatial alignment of star-forming galaxies (SFGs) and quiescent galaxies (QGs) in clusters. 
The results reveal that, in both brightest central galaxy (BCG) alignment (BA) and satellite alignment (SA), the alignment signals of SFGs and QGs exhibit opposite trends in several aspects. 
For the BA signal, SFGs show stronger alignment in cluster outskirts, whereas QGs exhibit stronger alignment in inner regions. 
For the SA signal, both SFGs and QGs exhibit nearly isotropic signals in outer regions, while in inner regions they show opposite signals. 
In addition, the BA signal depends on cluster richness and QG fraction, and is more pronounced in mature clusters. Specifically, QGs show a stronger BA signal overall, which increases with luminosity and axis ratio, whereas SFGs show a weaker BA signal with no significant dependence on these parameters. 
The BA signal also exhibits a strong dependence on satellite galaxy color, which is independent of morphology. 
We confirm this by analyzing the BA signal in two populations: red disk and blue elliptical galaxies.
However, neither BA nor SA shows a significant dependence on redshift at $z<0.5$. 
In summary, these findings reflect how galaxy transport along cosmic filaments drives cluster evolution: early-stage SFGs fall into clusters along filaments, and during cluster relaxation, their spatial orientations are reshaped by gravitational potentials and tidal forces, weakening BA signal and enhancing SA signal. 
In contrast, QGs, as early cluster members, exhibit a more pronounced anisotropic distribution and maintain relatively stable spatial orientations.

\end{abstract}


\section{Introduction}           
\label{sec:1}

Galaxy clusters, as the largest virialized systems in the universe, serve as natural laboratories for investigating galaxy evolution and tracing the cosmic large-scale structures. 
Clusters can be detected through multiple observational approaches. 
Besides optical, near-infrared, and gravitational lensing observations, the hot ionized intracluster medium (ICM) of clusters can be probed via X-ray emissions or radio observations of the Sunyaev-Zel'dovich (SZ) effect (\citealt{1971ApJ...165L..27G,1972CoASP...4..173S,1984ApJ...286..186F,2010PASJ...62..811O}). 
Meanwhile, using optical observations of the spatial distribution of member galaxies to constrain models of cluster dynamics significantly contributes to our understanding of gravity, dark matter, and the large-scale structure of the universe (e.g., \citealt{2006MNRAS.369.1293Y,2009ApJ...694..214O,2016MNRAS.463..222H,2018MNRAS.474.4772H,2025ApJ...983..122R,2025ApJ...987L..30W}).  

Studies of large-scale structure and galaxy clusters show that the observed anisotropy can be quantified by the alignment signals. 
At smaller scales (within the cluster), these signals manifest as differences in spatial orientation between satellite and central galaxies, as well as between the central galaxy and the cluster's orientation (e.g., \citealt{2009ApJ...694..214O,2016MNRAS.463..222H,2018MNRAS.474.4772H,2022MNRAS.516.3159Y}). 
On scales of $\sim$10 Mpc, the alignment signals observed between galaxy spin axes and cosmic web filaments provide important clues about the formation of clusters and the large-scale structure (e.g., \citealt{2013ApJ...779..160Z,2025ApJ...983..122R,2025ApJ...987L..30W}). 
Observations and simulations both show significant galaxy alignment signals out to about 70 $h^{-1}$ Mpc, but the alignment amplitude of dark matter halos in simulations is higher than that observed on all scales probed (\citealt{2013ApJ...770L..12L}).   
In addition, a non-intrinsic effect capable of distorting or enhancing galaxy tangential alignment signals arises from weak gravitational lensing. Such coherent lensing-induced alignments have been detected in observational data alongside the intrinsic alignments mentioned above (e.g., \citealt{2015PhR...558....1T,2015SSRv..193...67K}). 

As indicated by numerous observational results, a pronounced anisotropic distribution of member galaxies has been observed, particularly among redder satellite galaxies in closer orbits and in systems hosting redder and brighter central galaxies (e.g., \citealt{1969ArA.....5..305H,2006MNRAS.369.1293Y,2013ApJ...770L..12L,2015SSRv..193....1J,2015SSRv..193..139K,2022MNRAS.516.3159Y,2023MNRAS.525.4685S,2024ApJ...974...40L,2025A&A...700A.264D}). 
While early studies of alignment signals were contentious (e.g., \citealt{1968PASP...80..252S,1969ArA.....5..305H,1970IrAJ....9..211O}), the release of more comprehensive observational data leads to a growing consensus that these signals are critical probes for investigating the most pressing cosmological questions today (e.g., \citealt{1989ApJ...336...46W,1993ApJ...418..544V,1997ApJ...479..632S}). 
In clusters, the orientations of member galaxies, including satellite and central, are important indicators used to probe the alignment signal (\citealt{2006MNRAS.369.1293Y,2022MNRAS.516.3159Y,2025ApJ...992L..17W}).  
On the one hand, the ``central alignment” --- defined as the angle between the satellite-central galaxy connecting line and the central galaxy's major axis --- serves as a sensitive tracer of the dark matter halo\rq s gravitational field within the large-scale structure (e.g., \citealt{2008ApJ...675..146F,2010MNRAS.405.2023N,2018ApJ...859..115W}). 
This observed geometric alignment indicates that satellite orbital planes are preferentially aligned with the halo's major axis.  
On the other hand, the ``satellite alignment” --- the angular correlation between the satellite-central connecting vectors and the major axes of satellite galaxies --- provides a robust probe of environmental effects, particularly those induced by tidal torques and dynamical friction (e.g., \citealt{1994MNRAS.270..390C,2015MNRAS.450.2195S}). 
This alignment signal reflects how the intrinsic shapes and angular momenta of satellites are modified by external tidal fields and orbital energy loss. 

By detecting the two types of alignment angles defined earlier, we can obtain the strength of the alignment signal to analyze how these signals depend on intrinsic galaxy properties and environmental factors. 
For example, \cite{2006MNRAS.369.1293Y} analyze over 24,000 central-satellite pairs using data from the Sloan Digital Sky Survey (SDSS) and find a preferential alignment of satellites along the major axes of their central galaxies. 
They further show that systems with red central and red satellite galaxies exhibit stronger spatial alignment than those with blue central galaxies. 
In contrast, systems with blue centrals display a nearly isotropic satellite distribution.  
\cite{2016MNRAS.463..222H} employ the red-sequence Matched-filter Probabilistic Percolation (redMaPPer) algorithm to identify over 8,000 galaxy clusters, comprising nearly 100,000 galaxies, for a systematic investigation of the central alignment signal in cluster environments. 
Their analysis reveals that more elongated and richer clusters exhibit a stronger alignment signal with their central galaxies. 
More prominently, redder, brighter, and more centrally located satellite galaxies exhibit a pronounced tendency to align along the major axis of the central galaxy. 
Additionally, a follow-up study by \cite{2018MNRAS.474.4772H} find that the satellite alignment signal is statistically weaker and shows a certain degree of dependence on the choice of shape measurement method.  
 
In simulation work, these observational effects are further validated.  
Both gravitational (e.g., N-body) and hydrodynamical (e.g., IllustrisTNG) numerical simulations have revealed that within clusters, satellite galaxies align with the central galaxy, whereas on larger scales, galaxies tend to align with their host cosmic filaments (e.g., \citealt{2015SSRv..193...67K,2018MNRAS.481..414G,2020MNRAS.495.2436R,2020MNRAS.495.3002K,2024ApJ...974...40L}). 
More advanced mock observation techniques are applied to hydrodynamic numerical simulations to investigate alignment signals and their dependencies (e.g., \citealt{2020MNRAS.495.3002K,2024ApJ...974...40L,2025ApJ...992..157G,2025ApJ...992...41M} ). 
\cite{2020MNRAS.495.3002K} use hydrodynamical simulations to reproduce radial alignment and to demonstrate that the alignment signals of galaxies depend on neither stellar mass nor the dynamical state of their host clusters. 
Using TNG100 (\citealt{2018MNRAS.473.4077P}) data, \cite{2024ApJ...974...40L} investigate how the alignment signal depends on galaxy metallicity, as well as on the mass, age, and host halo mass of the brightest group galaxy (BGG). 
They find positive correlations among these parameters, which are consistent with observational results.    
While simulations employing different methodologies may yield varying outcomes, studies that integrate both observations and simulations remain essential for advancing our understanding of galaxy and cluster formation and evolution.  
    
This work employs data from the public Data Release 9 (DR9) of the Dark Energy Spectroscopic Instrument (DESI) Legacy Imaging Surveys (\citealt{2019AJ....157..168D}) to investigate the alignment between the orientations of satellite galaxies and the brightest central galaxies (BCGs) in clusters.   
Employing the Clustering by Fast Search and Find of Density Peaks (CFSFDP) algorithm, \cite{2021ApJS..253...56Z} identify 540,432 galaxy clusters at $z \lesssim 1$ from the photometric redshift catalog derived from the DESI Legacy Imaging Survey. 
Using this large sample of galaxy clusters, \citet{2024ApJ...961..253C} construct an environmental sample, which includes three density subsamples corresponding to field, cluster, and BCG environments, and classify galaxies by color ($g-z$) and specific star formation rate (sSFR) to study environmental effects. 
For a more detailed introduction to the DESI-based catalog, please refer to our series of articles (e.g., \citealt{2021ApJS..253...56Z,2024ApJ...961..253C,2025MNRAS.544.3350L}). 
Based on prior work in galaxy cluster research, this study focuses on the examination of the alignment between different types of galaxies and their central BCGs.     
        
This paper is organized as follows. 
In Section \ref{sec:2}, we describe the detailed processes for the construction of the cluster sample and the calculation of the key parameters.    
In Section~\ref{sec:3}, we analyze the alignment signal detected between quiescent and star-forming galaxy populations. 
Section~\ref{sec:4} discuss the dependence of alignment signals on different physical parameters of galaxies and on the cluster environment. 
Section~\ref{sec:5} presents the summary and conclusions of our work. 
Throughout this paper, we adopt the following cosmological parameters: $\Omega_m$ = 0.30, $\Omega_{\Lambda}$ = 0.70, and $\rm H_0$ = 70 km $\rm s^{-1}$ $\rm Mpc^{-1}$

\section{Data}
\label{sec:2}

\subsection{Galaxy Cluster Sample Construction}
\label{subsec:2.1}

\subsubsection{Overview of DESI imaging surveys}
\label{subsec:2.1.1}

The DESI Legacy Imaging Surveys are among the surveys with the widest sky coverage and the most comprehensive galaxy samples.  
Integrating data from its three constituent optical projects (BASS, MzLS, DECaLS) with those from DES and WISE, the surveys cover approximately half of the celestial sphere ($\sim$ 20,000 square degrees), provide imaging in five photometric bands ($g$, $r$, $z$, $W1$, $W2$), and catalog over 300 million galaxies (e.g., \citealt{2010AJ....140.1868W,2016arXiv161100036D,2016MNRAS.460.1270D,2019AJ....157..168D}). 

To enhance the utility of WISE mid-IR data, the Legacy Surveys employ a forced photometry approach implemented via the $Tractor$ algorithm (\citealt{2016ascl.soft04008L}) to construct galaxy catalogs. 
This approach performs unified modeling of multi-resolution optical and infrared data, enabling the detection of fainter sources than traditional methods and the deblending of objects that are confused in WISE images but resolved optically, while maintaining photometric consistency across bands. 
The $Tractor$ catalogs include fluxes, positions, and shape parameters, along with morphological quantities for distinguishing extended sources from point sources (\citealt{2019AJ....157..168D}). 

A key improvement in DR9 is the adoption of extended point-spread function (PSF) modeling (PSFEx; \citealt{2011ASPC..442..435B}) for stellar wing subtraction, together with a discrete six-type morphological classification scheme.  
This system provides rigorous, model-driven morphology measurements through $\chi^2$ minimization across multiple bands, enabling precise distinction between point sources (``PSF"), round exponential galaxies with a variable radius (``REX"), de Vaucouleurs (``DEV") profiles (elliptical galaxies), exponential (``EXP") profiles (spiral galaxies), and S\'{e}rsic (``SER") profiles. 
A source is retained if the better of PSF or REX improves the penalized $\chi^2$ by $\geq 25$. 
For retained sources, an extended profile (DEV/EXP) is adopted if it further improves the penalized $\chi^2$ by $\geq 9$ ($\approx 3\sigma$) over PSF/REX, corresponding to $\geq 5.8\sigma$ detections. 
A S\'{e}rsic profile is assigned only if it improves the penalized $\chi^2$ by another $\geq 9$ over DEV/EXP, corresponding to $\geq 6.5\sigma$ detections.  
\setcounter{footnote}{0} 
Further details are available on the official DESI \href{https://www.legacysurvey.org/dr10/catalogs/}{website}\footnote{\href{https://www.legacysurvey.org/dr10/catalogs/}{https://www.legacysurvey.org/dr10/catalogs/}}. 

\subsubsection{Cluster Sample}
\label{subsec:2.1.2}

Using the $Tractor$ catalog, \cite{2021ApJS..253...56Z} construct a large sample of galaxy clusters with $N_{\rm 1~Mpc} > 10$ (i.e., clusters with more than 10 satellite galaxies within 1 Mpc of the BCG) by employing the CFSFDP algorithm. 
The final sample includes over 40 million member galaxies, all at $z < 1$ and with $r < 23$ mag (AB). 
Target clusters are identified when the local galaxy density within a projected radius of $R < 0.5~\mathrm{Mpc}$ (in the redshift slice $z \pm 0.04~(1+z)$) exceeds the background density within the same slice at $R > 1~\mathrm{Mpc}$ by a factor of 4. 
Once the central galaxy (the local density peak) is identified, the cluster member galaxies are iteratively determined by selecting all galaxies with a distance of less than 0.5 Mpc to their nearest higher-density neighbor, and subsequently tracing these connections back to the same density peak. 
For further details regarding the galaxy cluster identification procedure, we refer readers to \cite{2020PASP..132b4101G} and \cite{2021ApJS..253...56Z}. 

The photometric redshift information, with an accuracy of $\sigma_z \approx 0.017$, can be found in \cite{2019ApJS..242....8Z,2022RAA....22f5001Z}. 
However, in this work we set the upper redshift limit to 0.5, as the physical size of a typical PSF in the $r$ band ($\sim 0.''8$) corresponds to $\sim 5$ kpc at $z = 0.5$.  
To maintain galaxy sample reliability, we consistently apply the quality cuts defined in \cite{2024ApJ...961..253C}:
\begin{itemize}
    \item Photometric redshift accuracy: $\sigma_{z_{\rm photo}}\leq 0.05$ $(1+z_{\rm phot}) $; 
    \item Size ($R_{\rm e}$) measurement quality: $R_{\rm e}/\sigma_{R_{\rm e}} > 5$; 
    \item Stellar mass range: $10^6 < M_*/M_{\odot} < 10^{13}$, with logarithmic mass uncertainty $|\Delta \log(M_*/M_{\odot})| < 0.4$; 
    \item Absolute magnitude cut: $-25 < M_r < -16$; 
    \item Apparent magnitude limit: $r < 22$ (to exclude faint sources). 
\end{itemize}

We select clusters with member galaxy richness $N_{\text{1 Mpc}} \geq 20$ to ensure both observational completeness of the cluster sample and statistical robustness of the results (\citealt{2024ApJ...961..253C}). 
These clusters constitute the final sample for statistical analysis. 
Additionally, to study the alignment signals between satellites and the BCG, we refine our cluster sample using the following strategy, which ensures the reliability and completeness of member galaxies. 

First, we define the normalized radial distance as $R/R_{200}$, where $R$ is the projected distance between the BCG and satellite galaxies. 
$R_{200}$ is the virial radius of the halo, converted from $M_{500}$ (defined as 500 times the critical density) using a mass definition transformation and then empirically corrected following \cite{2025ApJ...990..171C}. 
This normalized distance is then used to quantify the separation of each satellite from the BCG.  
Using the above distance parameters, we divide satellite galaxies into inner and outer members using a boundary at $0.5R_{200}$. 
Specifically, galaxies within $0.5R_{200}$ are classified as inner members, and those between $0.5R_{200}$ and $R_{200}$ as outer members. 
Any galaxy located beyond $R_{200}$ from the BCG is excluded from the cluster sample. 

Second, we also take into account the velocity dispersion ($v_{\text{los}}$) of the cluster as an important parameter for selecting member galaxies. 
Following \cite{2011Natur.477..567W}, we consider both the intrinsic dispersion of member galaxies relative to the BCG ($\pm 4000$ km s$^{-1}$) and the dispersion arising from photometric redshift errors of $\pm 7500$ km s$^{-1}$, which is estimated using the average velocity dispersion based on the photometric redshift accuracy of 0.017 reported by \cite{2019ApJS..242....8Z}. 
We therefore adopt $v_{\text{los}} < 8500$ km s$^{-1}$ as one of the selection criteria for member galaxies. 

The large number of galaxy samples allows us to adopt stricter criteria to select more reliable samples of star-forming galaxies (SFGs) and quiescent galaxies (QGs). 
The specific star formation rate threshold of $\log\,\mathrm{sSFR} = -11$ serves as an important criterion for distinguishing between SFGs and QGs (\citealt{2012MNRAS.424..232W}). 
On this basis, we also adopt the galaxy color $(g-z)$ as an additional criterion. 
Ultimately, galaxies with $-50 < \log\,\mathrm{sSFR} < -11$ and red colors are classified as QGs, while those with $-11 < \log\,\mathrm{sSFR} < -7$ and blue colors are classified as SFGs. See \cite{2024ApJ...961..253C,2025ApJ...995..103C} for details on the classification criteria.

\begin{figure}[ht!]  
    \includegraphics[width=0.45\textwidth]{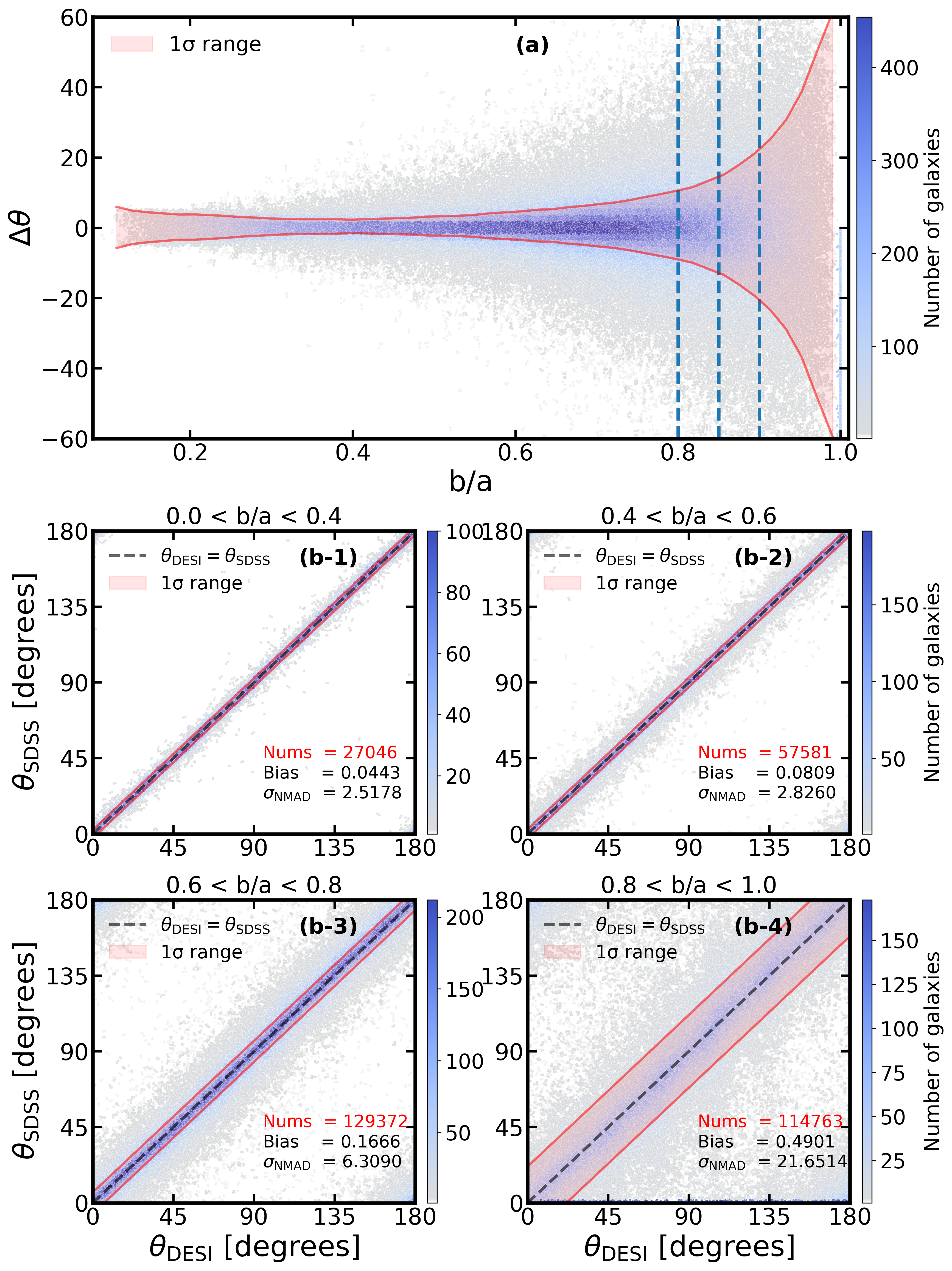}
    \centering 
    \caption{Comparison of position angle accuracy: 
        (a) shows the distribution of $\Delta\theta$ as a function of $b/a$, with vertical dotted lines marking $b/a = 0.8$, 0.85, and 0.9; 
        (b) displays the comparison of position angles for different $b/a$ bins, where the x-axis shows the DESI-measured position angle and the y-axis shows the SDSS-measured value. 
        The red label ``Nums" indicates the total sample count in each bin, ``Bias" denotes the fraction of galaxies with $|\theta_{\rm DESI} - \theta_{\rm SDSS}| > 15^\circ$, and $\sigma_{\mathrm{\scriptscriptstyle NMAD}}$ quantifies the median absolute deviation.
    \label{fig:2.1}}
\end{figure}  

\subsection{Key Parameters for Quantifying Alignment Signals}
\label{subsec:2.2}

\subsubsection{Orientation and Ellipticity}
\label{subsec:2.2.1}

In celestial coordinates, the position angle $\theta$ of a galaxy is defined, within $0^\circ \leq \theta < 180^\circ$, as the angle measured counterclockwise from the celestial north direction to the major axis on the tangent-plane projection of the sky.  
All shape parameters are derived from the ellipticity components ($\epsilon_1$, $\epsilon_2$) provided by the \textit{Tractor} catalog, where $\epsilon_1$ and $\epsilon_2$ are the two components of the complex ellipticity $\epsilon = \epsilon_1 + i\epsilon_2$. 
The key parameters and their interrelations are defined by the following formulas:
\begin{gather}
    \theta = \frac{1}{2} \arctan\!\left(\frac{\epsilon_2}{\epsilon_1}\right), \tag{1} \\
    |\epsilon| = \sqrt{\epsilon_1^2 + \epsilon_2^2}, \tag{2} \\
    \frac{b}{a} = \frac{1 - |\epsilon|}{1 + |\epsilon|}, \tag{3} \\
    e_c \equiv\sqrt{1 - \left(\frac{b}{a}\right)^2}, \tag{4}\\
    e_l \equiv 1-\frac{b}{a}. \tag{5}
\end{gather}
Here, $|\epsilon|$ is the magnitude of the complex ellipticity, $b/a$ is the axis ratio, $e_c$ is the geometric eccentricity, and $e_l$ is the linear ellipticity.  

To ensure the reliability of the galaxy position angle measurements, we adopt two approaches. 
First, we remove samples with large errors by applying a cut on the position angle error ($\theta_{\text{err}} < 10^\circ$). 
Second, we assess the reliability of the measurements by a direct comparison between the DESI and SDSS catalogs, using a matched sample of 328,762 galaxies constructed based on their RA, Dec, and redshift.  
In the matched SDSS catalog, the galaxy position angle is determined according to the criterion of ${\it fracDeV\_r} > 0.5$ (\citealt{2008MNRAS.388.1321P}): it is taken from the de Vaucouleurs model if ${\it fracDeV\_r} > 0.5$, and from the exponential model if ${\it fracDeV\_r} \le 0.5$.  
For each galaxy in this common sample, the position angle difference $\Delta\theta$ between the two surveys is calculated. 
The distribution of $\Delta\theta$ as a function of axis ratio ($b/a$) is presented in Figure~\hyperref[fig:2.1]{\ref*{fig:2.1}(a)}. It shows that the dispersion of $\Delta\theta$ increases sharply for nearly circular galaxies ($b/a > 0.8$). 
Figure~\hyperref[fig:2.1]{\ref*{fig:2.1}(b)} directly compares the position angles from the two surveys. 
We divide the data into four bins based on the axis ratio and analyze the dispersion of position angle measurements within each bin. 
Our results show that the discrepancy in position angles, quantified by $\sigma_{\mathrm{\scriptscriptstyle NMAD}}$, increases significantly when the axis ratio exceeds 0.8. 
As shown in Figure~\hyperref[fig:2.1]{\ref*{fig:2.1}(b-3)}, the $\sigma_{\mathrm{\scriptscriptstyle NMAD}}$ is $\sim$6.3, while in Figure~\hyperref[fig:2.1]{\ref*{fig:2.1}(b-4)}, it reaches $\sim$21.7.
Taking these factors into consideration, we apply an axis ratio cut of $b/a < 0.8$ as a criterion to minimize the impact of morphological measurement biases. 

The criterion of $b/a < 0.8$ also excludes all galaxies classified as ``REX" in the $Tractor$ catalog, since their position angles cannot be measured. 
This yields a final sample of 1,515,443 galaxies (81,815 BCGs and 1,433,628 satellite galaxies), within which we identify 934,731 QGs and 580,712 SFGs. 

\begin{figure}[ht!]  
    \includegraphics[width=0.45\textwidth]{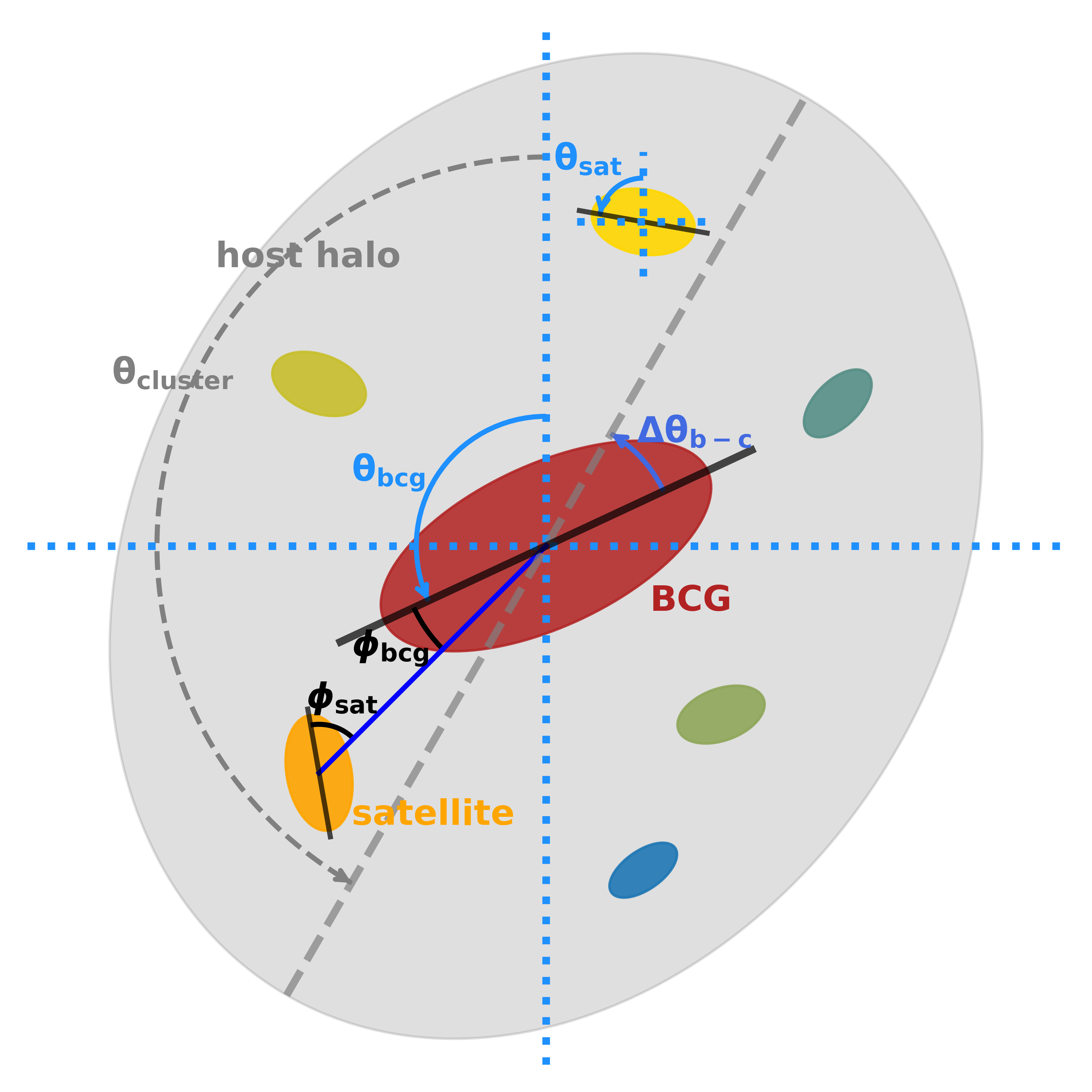}
    \centering 
    \caption{Illustration of the orientation and alignment angles of galaxies in a cluster. The position angles of the BCG, satellite galaxies, and the cluster are denoted by $\theta_{\rm bcg}$, $\theta_{\rm sat}$, and $\theta_{\rm cluster}$, respectively. The BCG alignment angle and the satellite alignment angle are denoted as $\phi_{\rm bcg}$ and $\phi_{\rm sat}$, respectively.   
    \label{fig:2.2}}
\end{figure}  

\subsubsection{Alignment Angles and Signals}
\label{subsec:2.2.2} 

To quantify the alignment signal, we use two alignment angles analogous to those defined in \cite{2016MNRAS.463..222H,2018MNRAS.474.4772H}. 
These two angles are used to quantify the degree of deviation between a satellite galaxy's major axis and that of its host system's central BCG. 
As illustrated in Figure~\ref{fig:2.2}, the BCG alignment (BA) angle ($\phi_{\rm bcg}$) is defined as the angle between the BCG's major axis and the vector from the BCG to a satellite galaxy. 
Conversely, the satellite alignment (SA) angle ($\phi_{\rm sat}$) is the angle between the satellite galaxy's own major axis and the same BCG-to-member vector. Both angles are confined to $[0^\circ, 90^\circ]$.   

The distribution of $\phi_{\rm bcg}$ values characterizes the spatial distribution trend of satellite galaxies relative to the BCG's major axis within a cluster. 
We compute the mean value of the BA angle, denoted as \(\langle \phi_{\rm bcg} \rangle\), for a given subsample to characterize the behavior of the BA angle distribution within that subsample. 
The difference \(45^\circ - \langle \phi_{\rm bcg} \rangle \) is defined as the BA signal. 
A positive difference indicates a radial alignment signal, meaning that the satellite galaxies are preferentially distributed along the major axis of the BCG. 
A negative difference indicates a tangential alignment signal, meaning that the satellite galaxies are preferentially distributed along the minor axis of the BCG. 
The BA signal can trace the anisotropic gravitational field of the dark matter halo and environmental effects embedded in large-scale structures. 

Similarly, the SA signal, defined as \(45^\circ - \langle \phi_{\rm sat} \rangle \), serves as a direct probe of the halo's gravitational potential, tidal torques, and the dynamical state of the satellite galaxies. 
A negative SA signal indicates that the major axes of satellite galaxies tend to be oriented perpendicular to the direction toward the cluster center, while a positive SA signal indicates that the major axes of satellite galaxies tend to point toward the cluster center.

\section{Alignment Signal Analysis} 
\label{sec:3}  

Studies demonstrate that the position angle of the BCG exhibits strong alignment with the orientation of the cluster halo (\citealt{2022MNRAS.516.3159Y}), implying that the spatial orientation of the BCG serves as an excellent tracer of the gravitational field of cluster halos. 
By analyzing the alignment between satellite galaxies and the BCG, we can connect the evolution of galaxies and galaxy clusters to the large-scale structure of the universe (\citealt{2020ApJ...900..129W}). 
The alignment signal serves as an indicator of both the cluster's relaxation process and the anisotropic infall of surrounding galaxies and intergalactic medium material along cosmic filaments. 
This process further modulates the star formation and evolutionary history of satellite galaxies.

\subsection{Distribution of BCG Alignment Angle}
\label{subsec:3.1}

\begin{figure*}[ht!]  
    \includegraphics[width=0.95\textwidth]{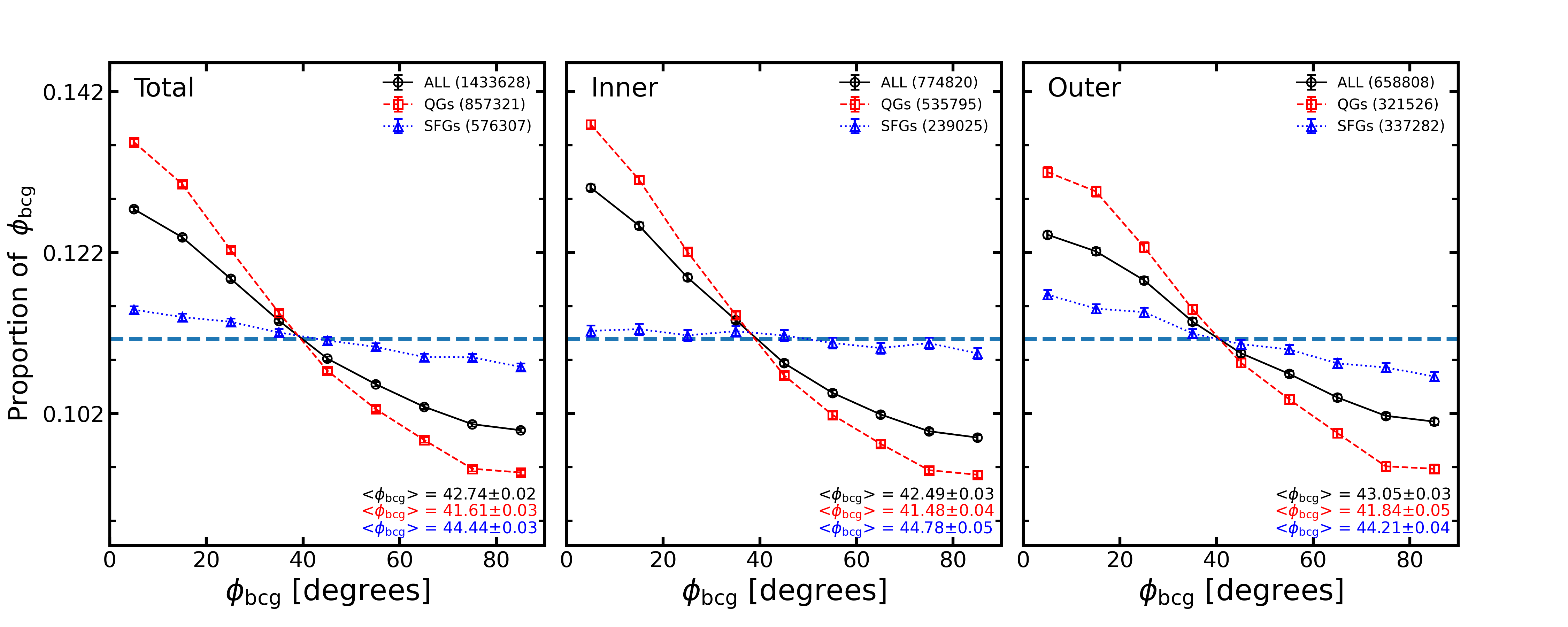}
    \centering 
    \caption{The distribution of $\phi_{\rm bcg}$ is plotted with the x-axis divided into 9 bins, and the y-axis showing the normalized galaxy count per bin (normalized by the total subsample size). The left, middle, and right subplots correspond to the distributions of the total member galaxy sample, the inner region cluster members, and the outer region cluster members, respectively. The black solid line connects circles with error bars, representing all member galaxies combined, while red squares (dashed line) and blue triangles (dotted line) indicate QGs and SFGs. The top-right legend displays the subsample sizes, and the bottom-right panel lists the mean $\langle \phi \rangle$ values for each galaxy type. The horizontal blue dashed line indicates the expected distribution under the assumption of isotropic alignment angles. The error bars represent the standard errors of the proportions, calculated as $\sqrt{p(1-p)/N}$, where $p$ is the proportion of galaxies in each bin and $N$ is the total sample size.   
    \label{fig:3.1}}
\end{figure*}  

We analyze the distribution of BA angle $\phi_{\rm bcg}$ for 1,433,628 satellite galaxies covering more than 140,000 clusters, with the results shown in Figure~\ref{fig:3.1}.  
Compared to SFGs, the BA angle distribution of QGs exhibits a significant alignment trend in both the inner and outer regions of clusters. 
The mean BA angle $\langle \phi_{\rm bcg} \rangle$ for QGs exhibits a small but significant radial gradient, increasing from $41.48^{\circ} \pm 0.04^{\circ}$ in the inner regions of the cluster to $41.84^{\circ} \pm 0.05^{\circ}$ in the outer regions. 
The uncertainties of the mean values are given as the standard errors of the mean, $\sigma/\sqrt{N}$, where $\sigma$ is the standard deviation of the $\phi$ distribution. 
In general, a strong positive BA signal (i.e., $\langle \phi_{\rm bcg} \rangle < 45^\circ$) may result from both the large-scale cosmic web structure and the dynamical evolutionary history of galaxy clusters.  

As a comparison, SFGs show a relatively weak positive BA signal overall, compared to QGs.   
Interestingly, despite their weak alignment signals, SFGs exhibit a trend opposite to that of QGs, whose BA signal is stronger in the inner regions than in the outer regions. 
Specifically, in cluster outskirts the mean BA angle is $\langle \phi_{\rm bcg} \rangle = 44.21^{\circ} \pm 0.04^{\circ}$, yet this signal is significantly stronger than in inner regions ($44.78^{\circ} \pm 0.05^{\circ}$). 
Understanding this requires considering the distinct formation and evolutionary histories of SFGs and QGs within clusters. 
However, these results confirm the finding of \cite{2016MNRAS.463..222H} that red-sequence satellite galaxies with redder colors and smaller distances to the central galaxy exhibit stronger BA signals.  
It is worth noting that \cite{2016MNRAS.463..222H} also find that different morphological measurement methods can marginally affect the results; nevertheless, these differences remain within the margin of error of the BA signal, and thus do not affect our conclusions. 

Rather than discussing the effect of morphological differences on alignment signals, this work focuses on how the alignment signals of member galaxies vary with their star formation histories, aiming to further elucidate the relationship between galaxy alignment and cluster relaxation.   
The relaxation process of galaxy clusters results from the co-evolution of internal dynamical processes (e.g., galaxy mergers, tidal interactions, and dynamical friction) and the external large-scale cosmic web structure.  
This process helps us understand the formation and evolutionary history of galaxy clusters. 
It should be emphasized that our study shows the dependence of the BA signal on galaxy color is much stronger than that on galaxy morphology (Section~\ref{subsec:4.4} ), and the contrasting alignment signals of QGs and SFGs in Figure~\ref{fig:3.1} provide a unique perspective. These aspects are further discussed in Section~\ref{sec:4}.

\subsection{Distribution of Satellite Galaxy Alignment Angle}
\label{subsec:3.2}

Compared to the attention given to BCG alignment, satellite galaxy alignment has been relatively less studied, a situation largely attributable to the difficulty of detecting its alignment signal.  
As expected, our result exhibits a weaker SA signal than the BA signal, consistent with \cite{2018MNRAS.474.4772H}. 
However, by comparing the QG and SFG samples, we find a relatively pronounced anti-correlated alignment signal in the inner regions of clusters.   
In any case, during cluster relaxation and the accretion of intergalactic material along cosmic filaments, the spatial orientation of satellite galaxies may also contribute to our understanding of the virialization process of galaxy clusters.  

\begin{figure*}[ht!]  
    \includegraphics[width=0.95\textwidth]{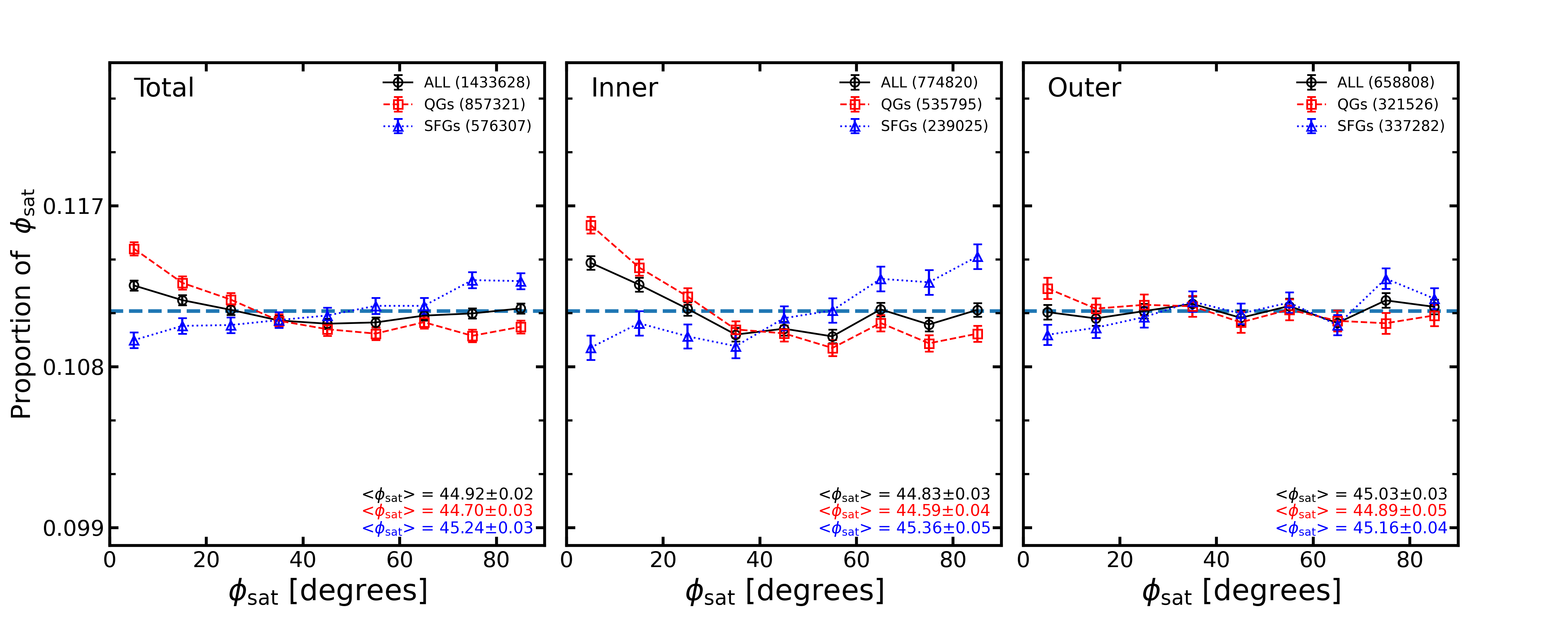}
    \centering 
    \caption{ The distribution of satellite galaxy alignment angle $\phi_{\rm sat}$ is plotted with the x-axis divided into 9 bins, and the y-axis showing the normalized galaxy count per bin (normalized by the total subsample size). The legend corresponds to Figure~\ref{fig:3.1}.
    \label{fig:3.2}}
\end{figure*}

As illustrated in Figure~\ref{fig:3.2}, our study reveals several unique observational results.  
First, in contrast to the approximately isotropic distribution observed in the outer regions of clusters, satellite galaxies in the inner regions exhibit a detectable SA signal. 
Second, among the satellite galaxies in the inner regions, SFGs and QGs show opposite alignment signals: the major axes of QGs tend to point toward the cluster center, while those of SFGs tend to be oriented along the tangential direction  (perpendicular to the cluster-centric radial direction). 
Specifically, the value of $\langle \phi_{\rm sat} \rangle$ in inner regions reaches $44.59^{\circ} \pm 0.04^{\circ}$ for QGs, which show a positive SA signal. 
In contrast, SFGs show a negative SA signal in inner regions, with $\langle \phi_{\rm sat} \rangle = 45.36^{\circ} \pm 0.05^{\circ}$, suggesting a distinct virialization process for SFGs compared to QGs, though an observation effect cannot be ruled out. 
 
The opposite SA signals exhibited by QGs and SFGs in the inner regions can be understood from three perspectives.  
First, we cannot rule out the contribution of weak gravitational lensing shear to the SA signal of SFGs. 
Compared to QGs, background SFGs are more prone to being misidentified as cluster members because of their larger photometric redshift errors (\citealt{2010ApJ...725..794Q}). 
Second, during the relaxation process of member galaxies evolving toward virialization, young SFGs experience stronger dynamical friction and ram pressure stripping due to their rich gas content, leading to faster loss of orbital angular momentum and changes in their orbits; meanwhile, during infall toward the cluster center, the combination of incomplete initial angular momentum dissipation and tidal torques from the cluster results in a weak tangential SA signal in these galaxies (\citealt{2010ApJ...721..939P}). 
Furthermore, when passing through the pericenter of their cluster-centric orbits, SFGs cannot instantaneously adjust their orientation to the rapidly varying tidal field due to inertia, and this lag increases the SA angle (\citealt{2008ApJ...672..825P}). 

We must emphasize that, as shown in \cite{2018MNRAS.474.4772H}, the SA signal \(45^\circ - \langle \phi_{\rm sat} \rangle \) shows some dependence on the galaxy morphology measurement method. 
In addition, by comparing the SA signals of foreground and background galaxies, \cite{2018MNRAS.474.4772H} reveal the contribution of weak gravitational lensing. 
However, given the limited precision of photometric redshifts, we are unable to provide a high-precision analysis of the weak lensing shear effect, and differences in sample selection may also contribute to the observed shear effect.  
In summary, we suggest that the opposing SA signals observed in the inner regions of clusters result from a combination of multiple factors. 
However, given the potential contribution of the weak gravitational lensing effect, we will examine whether it contributes at all in Section~\ref{sec:4}, using a sample with higher photometric redshift precision.

\section{Analysis and Discussion}
\label{sec:4}

\subsection{ Reliability of the Alignment Signal }
\label{subsec:4.1}

The results presented in Section~\ref{sec:3} show that there is a significant difference in the signal strengths of BA and SA, and both exhibit a clear color dependence. 
Before discussing the signal strengths of BA and SA, we further test the reliability of the alignment signal. 
We incorporate the first data release (DR1; \citealt{2026AJ....171..285D}) of the DESI spectroscopic survey and match cluster member galaxies with spectroscopic redshifts. 
This matching yields over 150,000 member galaxies, and we analyze their BA and SA signals. 
The results show that the mean BA angles for SFGs and QGs are \(44.41^{\circ} \pm 0.14^{\circ}\) and \(41.62^{\circ} \pm 0.08^{\circ}\), respectively; the mean SA angles for SFGs and QGs are \(45.19^{\circ} \pm 0.14^{\circ}\) and \(44.20^{\circ} \pm 0.08^{\circ}\), respectively. 
These results are generally consistent with those presented in Figures~\ref{fig:3.1} and~\ref{fig:3.2}, which show a positive BA signal for both SFGs and QGs, and a positive SA signal for QGs but a negative SA signal for SFGs.  

Results show that the SA signal is significantly weaker than the BA signal, partly due to the more obvious intrinsically random spatial orientations of satellite galaxies, and partly due to the challenges in measuring their morphologies. 
\cite{2018MNRAS.474.4772H} systematically analyzes the potential impact of galaxy morphology measurements on the alignment signal and points out that different morphology measurement methods have different sensitivities to isophotal twisting (\citealt{2010ApJ...721..939P}), with isophotal shape measurements showing the strongest signal.  
Shape measurement methods that concentrate on the inner structure of galaxies yield a weak, or even null, signal, whereas methods that emphasize the outer isophotes produce a stronger signal.  
Compared to the position angle measured by the isophotal method, the position angle measured using $Tractor$ in this paper represents an averaged, global orientation obtained by fitting the entire light distribution of a galaxy with a changeable parametric model. 
We cannot exclude the effect of differences in morphology measurements on the SA signal, but this effect is expected to be isotropic and to have only a limited impact on the signal. 
However, to minimize the potential impact of morphological measurement biases, in Section~\ref{subsec:2.2} we ensure the reliability of the sample by (i) excluding galaxies with large morphological uncertainties based on a comparison with SDSS, and (ii) retaining only galaxies with small position angle errors (e.g., $\theta_{\text{err}} < 10^\circ$).

\begin{figure*}[ht!]  
    \includegraphics[width=0.95\textwidth]{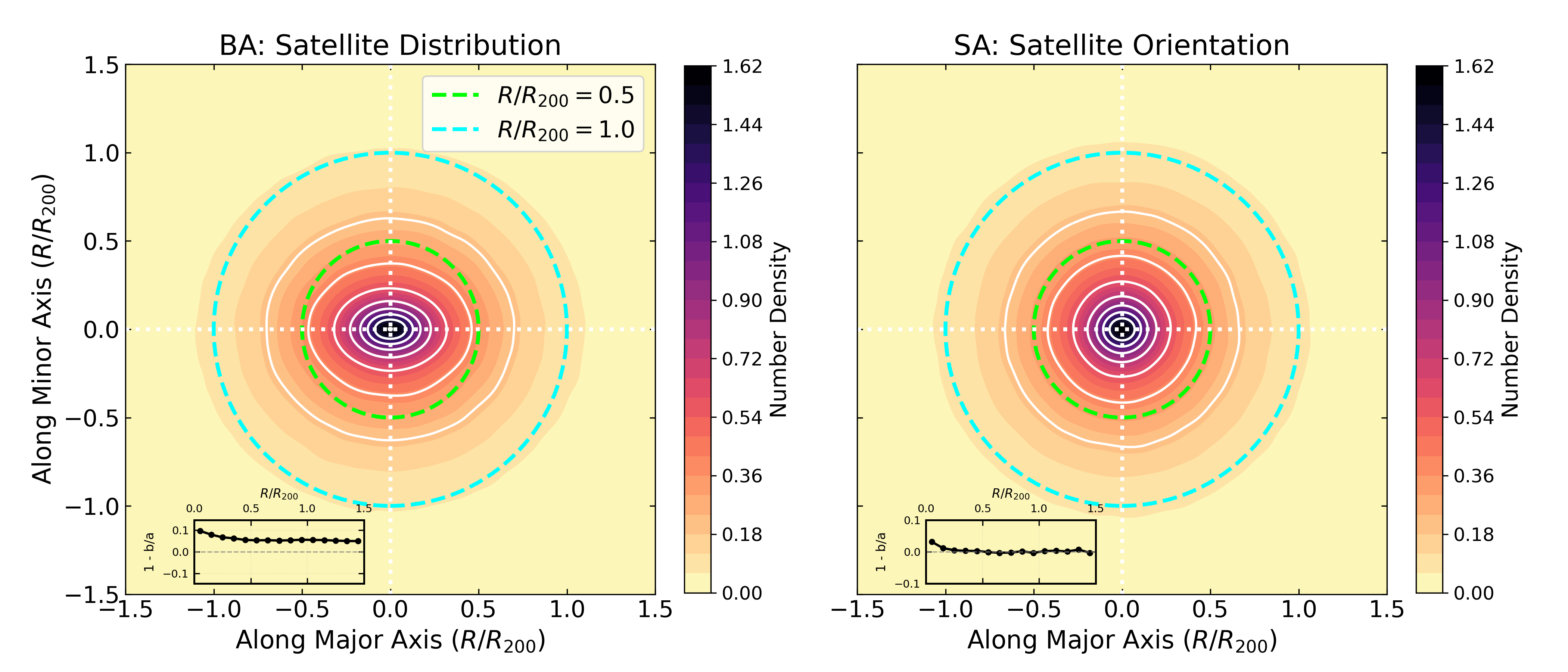}
    \centering 
    \caption{ Two-dimensional kernel density maps of satellite galaxies in the BA (left panel) and SA (right panel) coordinate systems. For the BA map, the coordinate frame of each cluster is rotated such that the major axis of the BCG is aligned with the horizontal direction. The stacked distribution therefore illustrates the projected spatial distribution of satellite galaxies relative to the BCG major axis. For the SA map, the coordinate frame of each satellite galaxy is rotated such that its own major axis is aligned with the horizontal direction, and the position of the BCG is measured in this frame, probing the radial alignment of satellite galaxies. Distances are normalized by the host cluster virial radius ($R_{200}$). The colour scale represents the two-dimensional kernel density estimate of the stacked distribution, while the white contours denote iso-density levels. The dashed green and cyan circles indicate projected radii of 0.5$R_{200}$ and 1.0$R_{200}$, respectively. The inset subplot with the black line shows the radial trend of shape ellipticity (1-$b/a$). Positive/negative values indicate elongation along the x/y-direction. 
    \label{fig:4.1}}
\end{figure*}  

\begin{figure*}[ht!]  
    \includegraphics[width=0.95\textwidth]{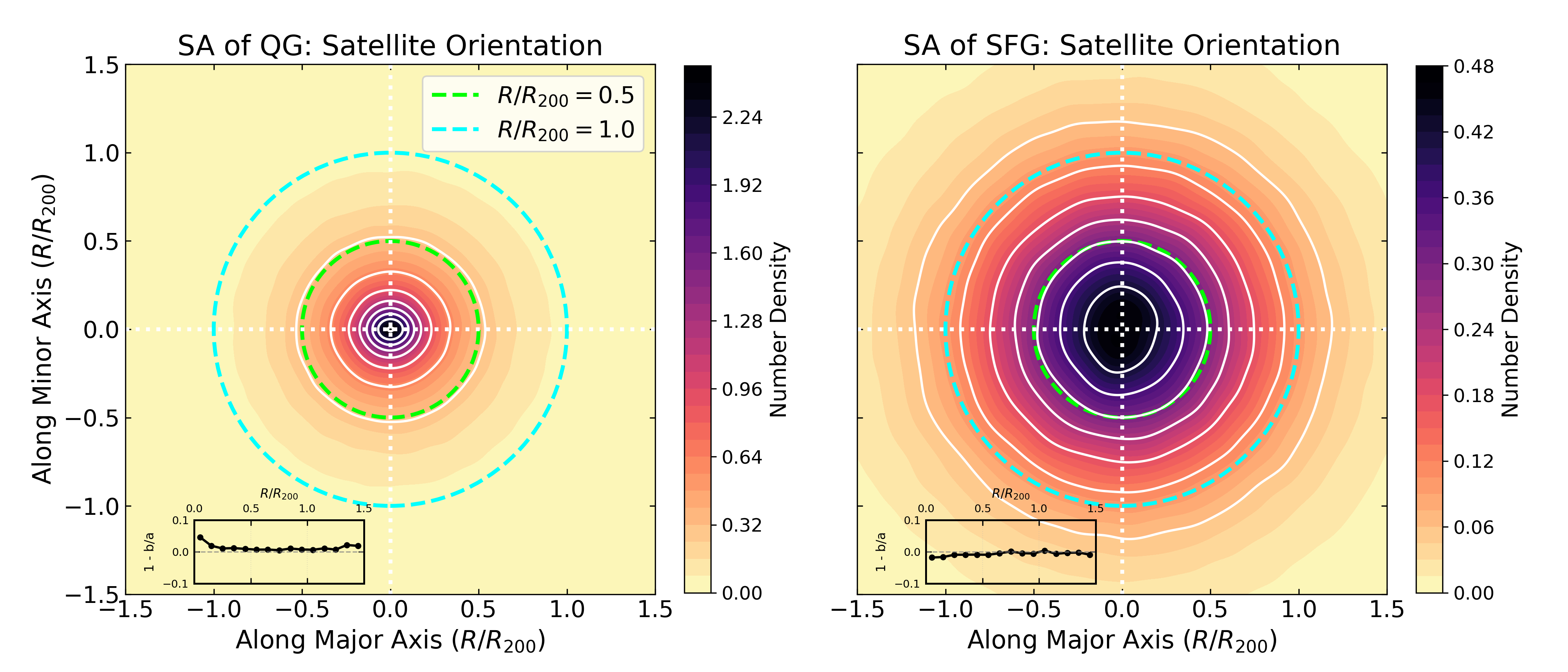}
    \centering 
    \caption{ Same as Figure~\ref{fig:4.1}, but for QGs (left panel) and SFGs (right panel) in the SA coordinate system. 
    \label{fig:4.2}}
\end{figure*}  

\subsection{ Anti-correlation of the Alignment Signal Between SFGs and QGs }
\label{subsec:4.2}

An anti-correlation trend between SFGs and QGs is found in both the BA and SA signals. 
For the BA signal, as shown in Figure~\ref{fig:3.1}, the QG alignment signal is stronger in the inner regions of the cluster, while the SFG alignment signal is stronger in the outer regions. 
This difference is likely closely related to the infall time of member galaxies into the cluster. 
QGs tend to be the original inhabitants of early clusters, with longer relaxation times. 
Their distribution properties result from long-term cluster relaxation, which causes them to align with the gravitational field distribution of the entire cluster. 
The BA signal is more pronounced in the inner regions, where early accreted galaxies have experienced prolonged exposure to the high-density environment, resulting in stronger tidal interactions and more noticeable alignment.  
In contrast, SFGs are mostly members that have migrated into the cluster later.  
Besides newly formed SFGs from subhalos within the cluster, there are also SFGs that have recently been accreted from cosmic large-scale structure filaments. 
These newly accreted SFGs often retain some of their filamentary properties, the most notable of which is a high degree of alignment between the cluster's major axis and the filament (\citealt{2007MNRAS.378.1531K}). 
SFGs accrete onto the galaxy cluster along filaments, retaining their primordial alignment signal, which is gradually disrupted by the cluster's gravitational field during infall toward the cluster center. 
However, this issue is discussed in more detail in Section~\ref{subsec:4.3}.  

For the anti-correlated SA signal, Figure~\ref{fig:3.2} shows that it is mainly observed in the inner regions of clusters, where SFGs and QGs exhibit opposite alignment signals. 
Specifically, the major axes of QGs tend to point toward the cluster center, whereas those of SFGs tend to be oriented tangentially to the cluster center. 
In the outer regions, the SA signals of both galaxy populations tend to be nearly isotropic, showing only a very weak signal. 
Our results indicate that, compared to the nearly isotropic distribution in the outer regions of galaxy clusters, the anisotropic distribution observed in the inner regions appears to be the result of multiple factors acting together. 
Understandably, under long-term tidal torques, the major axes of QGs tend to point toward the cluster center, particularly those in the inner cluster regions. 
However, to understand the tangential alignment of the major axes of SFGs, we can consider the following three aspects. 
First, the isophotal twisting effect can produce a tangential alignment signal in SFGs. 
Due to their rich gas content, SFGs in the inner regions of clusters are more susceptible to tidal forces, which in turn bias their position angles toward the tangential direction, as shown in Figure 6 of \cite{2010ApJ...721..939P}. 
Second, under the combined influence of gravity and tidal torquing, SFGs with shorter relaxation times and large angular momentum exhibit an increase in $\phi_{\rm sat}$ when they pass through the pericenter of their orbits around the cluster centroid, as illustrated in Figure 8 of \cite{2008ApJ...672..825P}. 
This is due to inertia, which prevents these SFGs from instantaneously adjusting their orientation to the rapidly varying tidal field when passing through the pericenter, and this lag leads to an increase in $\phi_{\rm sat}$. 
Finally, as discussed in Section~\ref{subsec:3.2}, we cannot rule out the contribution of gravitational lensing effects to the tangential alignment signal. 

To examine the contribution of the weak gravitational lensing effect to the SA signal of SFGs in the inner region of the cluster, we improve the precision of member galaxy redshifts to \(\sigma_{z_{\rm photo}}\leq 0.03\) \((1+z_{\rm phot})\) and measure the SA signal of SFGs. 
We find that in the inner region, \(\langle \phi_{\rm sat} \rangle = 45.24^{\circ} \pm 0.14^{\circ}\). 
Compared to \(\langle \phi_{\rm sat} \rangle = 45.36^{\circ} \pm 0.05^{\circ}\) in the original sample, the signal weakens slightly. 
Based on this result, we can confirm that the weak gravitational lensing effect contributes to the SA signal, i.e., the signal in the original sample includes shear contributions from background galaxies lensed by the cluster halo. 
This effect is more pronounced for SFGs than for QGs in the inner cluster regions, as their larger photometric redshift errors make background SFGs more prone to being misidentified as satellite galaxies.  
It should be emphasized that we also check the BA signal and the SA signal of QGs in this higher-precision sample and find no significant effect on these signals; therefore, this does not affect our main conclusions. 

To provide a more intuitive view of the spatial distribution and orientation of satellite galaxies within galaxy clusters, we present in Figure~\ref{fig:4.1} the stacked two-dimensional kernel density maps of satellite galaxies in the BA and SA coordinate systems. 
The left panel takes the BCG as the origin and projects satellite galaxy positions onto its major and minor axes to obtain the x and y coordinates; the right panel takes each satellite galaxy's own position as the origin and projects the BCG's position onto that satellite's major and minor axes.    
In both panels, the density peaks at the cluster centre and gradually declines towards larger radii. 
Additionally, as shown in the subplots of each figure, we quantify the radial trend of the shape ellipticity (1-$b/a$) by binning the point set radially and computing the covariance matrix of the point distribution within each bin, where positive values indicate elongation along the x-direction and negative values indicate elongation along the y-direction. 
The density contours in the BA map exhibit a clear elongation along the major-axis direction, deviating from the circular symmetry expected for an isotropic distribution. 
This feature indicates that satellite galaxies preferentially reside along the major axis of the BCG. 
In contrast, the SA map does not show a similarly pronounced trend, although a weak elongation along the major-axis direction can still be seen in the central region, likely driven by the contribution of QGs. 
Figure~\ref{fig:4.2} presents the stacked two-dimensional kernel density maps for QGs and SFGs in the SA coordinate system. 
QGs display an elongation along the major-axis direction in the SA coordinate system, indicating that their major axis is primarily aligned with the radial direction toward the BCG. 
In contrast, SFGs show a relatively stronger elongation along the minor-axis direction in the central region of the SA coordinate system, suggesting that their major axes tend to be oriented perpendicular to the cluster radial direction. 

In summary, relative to SFGs, QGs are members that fell into the cluster earlier, have longer relaxation times, sparser internal gas, and exhibit stronger spatial alignments.
Based on the results shown in Figure~\ref{fig:3.1}, we find that SFGs in the outer regions of clusters exhibit a significantly stronger BA signal than those in the inner regions. 
This finding serves as key evidence that SFGs are falling into the cluster along filaments. 
In contrast, the results shown in Figure~\ref{fig:3.2} indicate that the major axes of SFGs in the outer regions of clusters (i.e., the SA angle) exhibit a nearly isotropic distribution. 
This can be explained by their complex spin evolution history, which involves early laminar accretion perpendicular to filaments, followed by later spin realignment through mergers and anisotropic accretion along filaments (\citealt{2025ApJ...987L..30W}). 
When these SFGs fall into the inner regions of the cluster and pass pericenter, the transition of their orbital motion from radial to tangential subjects them to a rapidly varying tidal field, leading to a realignment of their major axes. 
Combined with the twisting effect of strong tidal interactions, this leads to a statistically weak tangential alignment signal for their major axes in the inner regions (\citealt{2008ApJ...672..825P}). 
These combined factors collectively explain the observed anti-correlation in the SA signal in the cluster inner regions. 

\subsection{Dependence of the BCG Alignment Signal }
\label{subsec:4.3}

\begin{figure*}[ht!]  
    \includegraphics[width=0.95\textwidth]{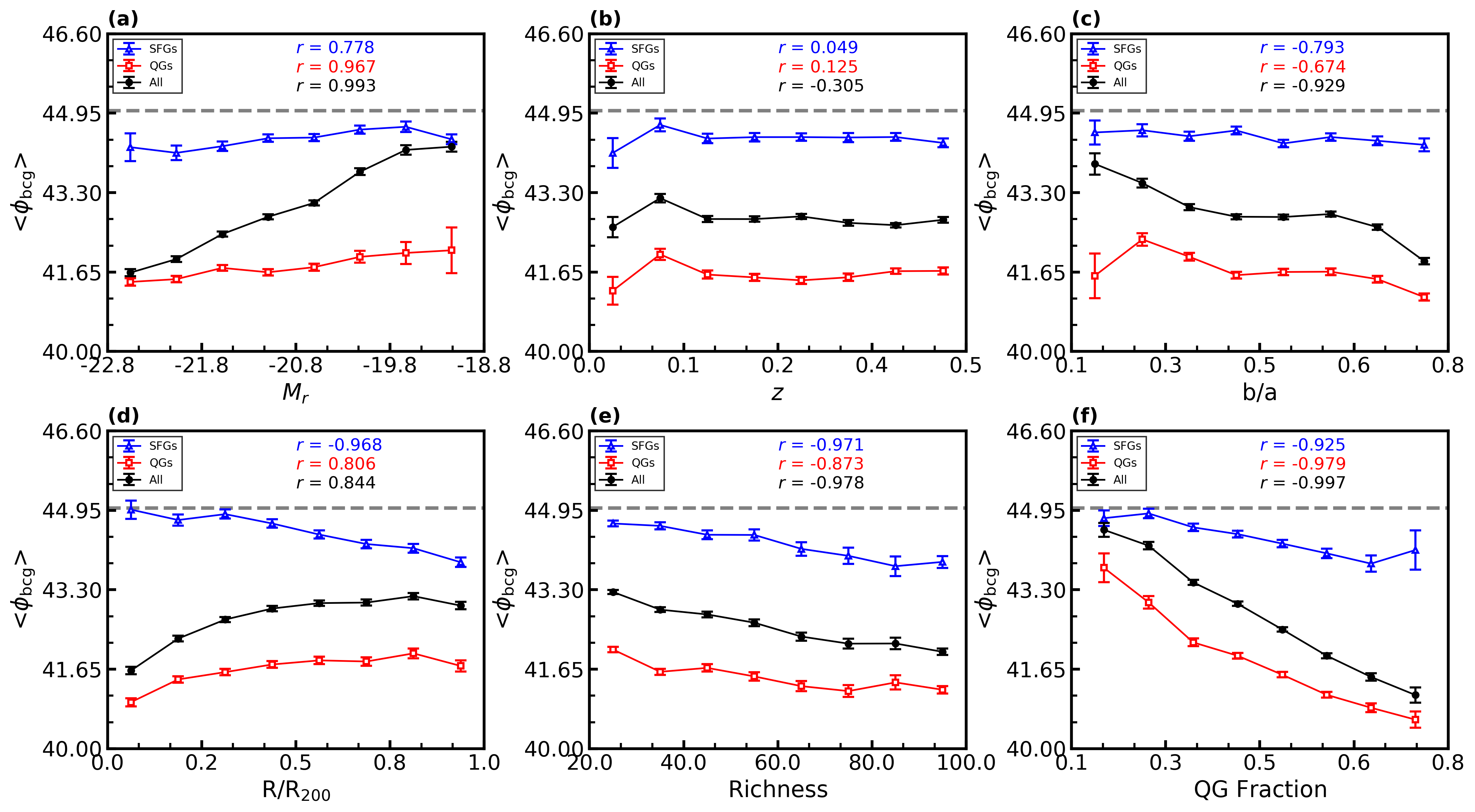}
    \centering 
    \caption{ Panels (a)$\sim$(f) present the mean values of BA angle ($\langle \phi_{\rm bcg} \rangle$) as functions of: absolute magnitude, redshift, axis ratio, and radial distance of satellite galaxies; and cluster richness and QG fraction. Blue hollow triangles, red hollow squares, and solid circles denote SFGs, QGs, and the total sample, respectively. The legends in the upper middle list their Pearson correlation coefficients ($r$). 
    \label{fig:4.3}}
\end{figure*}  

The dependence of the BA signal on the properties of the member galaxies themselves, as well as on the properties of the cluster, has been discussed in many studies. 
A broad consensus has been reached regarding its strong dependence on certain characteristics. 
For example, the BA signal exhibits a strong dependence on galaxy color, with red satellite galaxies or cluster systems dominated by red central galaxies showing a strong alignment signal (\citealt{2006MNRAS.369.1293Y,2007MNRAS.378.1531K,2016MNRAS.463..222H,2018ApJ...859..115W,2025A&A...700A.264D}). 
In addition, as shown in \cite{2016MNRAS.463..222H}, the BA signal exhibits a certain level of dependence on the ellipticities and absolute magnitudes of both central and satellite galaxies, as well as on the ellipticity and richness of the host clusters. 

Based on the characteristics of our sample, we examine the dependence on six parameters: the absolute magnitude ($M_r$), redshift ($z$), axis ratio ($b/a$), and radial distance ($R/R_{200}$) of satellite galaxies, the richness of the galaxy clusters, and the relaxation state of the clusters (traced by the fraction of QGs). 
As shown in Figure~\ref{fig:4.3}, taking the mean BCG alignment angle as a tracer of the alignment signal strength, we bin the six parameters individually to analyze the dependence of the alignment signal on each parameter. 
We consider an absolute Pearson correlation coefficient $|r| > 0.80$ to indicate a relatively strong linear correlation, and the key results are as follows.   

(\romannumeral 1) We find that, compared to SFGs, QGs exhibit a relatively strong dependence on $M_r$.  
Though the overall sample, indicated by the black solid circles in Figure~\hyperref[fig:4.3]{\ref{fig:4.3}(a)}, shows a very strong dependence on $M_r$, this dependence is not intrinsic to the individual populations but rather results from the difference in their sample sizes at the bright and faint ends. 
The proportion of QGs in the bright end is close to 75\%, while that in the faint end is about 11.5\%. 

(\romannumeral 2) Our results also confirm that $\langle \phi_{\rm bcg} \rangle$ shows no significant dependence on redshift (at $z<0.5$), as shown in Figure~\hyperref[fig:4.3]{\ref{fig:4.3}(b)}. 

(\romannumeral 3) For the galaxy axial ratio $b/a$, it is expected that a smaller value leads to a stronger BA signal, due to the stronger coupling with tidal torques. 
However, \cite{2006MNRAS.369.1293Y} find that higher ellipticity corresponds to a weaker alignment signal, which they attribute to the fact that galaxies with ellipticity greater than 0.6 are predominantly blue, late-type disc galaxies (edge-on galaxies). 
Extending their work, the present study focuses on the dependence of the BA signal on satellite galaxy morphology traced by $b/a$. 
The results are shown in Figure~\hyperref[fig:4.3]{\ref*{fig:4.3}(c)} and indicate that QGs with $b/a > 0.2$ tend to show a dependence, while SFGs show no significant dependence.  
However, the significant variation at $b/a < 0.2$ exhibited by the black solid line may be a result of the increasing fraction of SFGs among high-ellipticity galaxies (\citealt{2006MNRAS.369.1293Y}).   
We calculate the proportion of QGs in different $b/a$ intervals and find that it increases with increasing $b/a$. 
For $b/a < 0.2$, the proportion of SFGs is more than 80\%.  

(\romannumeral 4) Investigating how the BA signal depends on the radial distance $R/R_{200}$ from the cluster center, we find that SFGs and QGs exhibit opposite dependencies.  
As shown in Figure~\hyperref[fig:4.3]{\ref*{fig:4.3}(d)}, the BA signal strengthens with increasing distance from the cluster center for SFGs, but weakens with increasing distance for QGs. 
This is consistent with the observation in Figure~\ref{fig:3.1} that SFGs in the inner region show no significant alignment while those in the outer region exhibit a non-negligible alignment signal.  
However, it should be emphasized that QGs exhibit a significantly stronger alignment signal than SFGs overall. 
It is understandable that during the virialization process of clusters, QGs that have undergone long-term dynamical relaxation tend to be preferentially distributed along the major axis of the BCG. 
In contrast, SFGs, which are often dynamically younger or newly infalling into the cluster, exhibit a more random distribution. 
Nevertheless, the obvious radial dependence of SFGs, specifically that the BA signal becomes stronger farther from the cluster center, may arise from two factors. 
Part of the reason is that the anisotropic accretion during cluster growth manifests as preferential accretion of material from larger dark matter halos along the major axis (\citealt{2007MNRAS.378.1531K}). 
This process explains why SFGs in outer regions show an obvious BA signal. 
The other part arises from rearrangement effects: as SFGs fall toward the cluster center, the combined action of their initial angular momentum and the host halo's gravitational field causes them to be rearranged.  
 
(\romannumeral 5) Cluster richness is an important parameter for tracing the origin and evolutionary history of clusters.  
In richer clusters, the deeper gravitational potential wells lead to more efficient dynamical relaxation, stronger tidal interactions, and faster environmental quenching among member galaxies.  
This anisotropic environment further drives the preferential accretion of satellite galaxies along the BCG's major axis, resulting in stronger BA signals.  
Our work confirms this, as shown in Figure~\hyperref[fig:4.3]{\ref{fig:4.3}(e)}: both SFGs and QGs display a strong dependence on cluster richness.  
This result implies that the BA signal of member galaxies within clusters is a natural consequence of long-term dynamical relaxation.   

\begin{figure*}[ht!]  
    \includegraphics[width=0.95\textwidth]{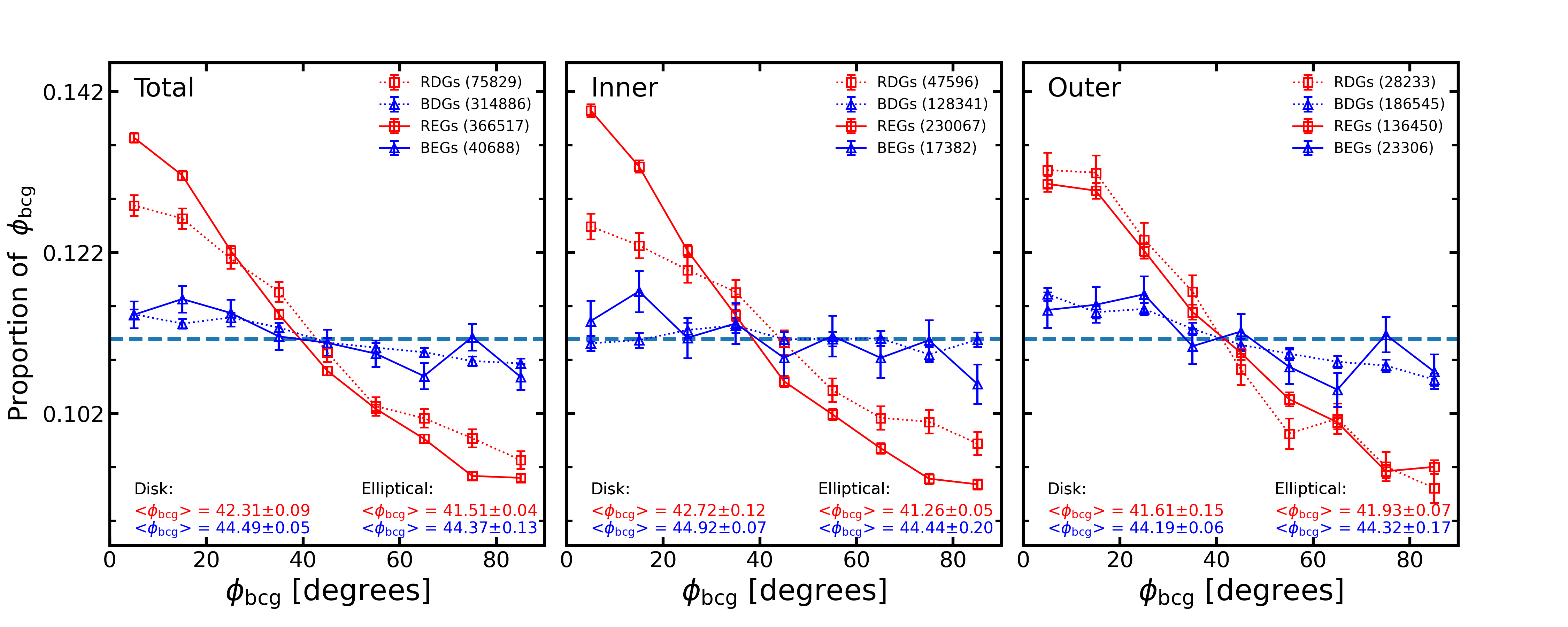}
    \centering 
    \caption{ The distribution of BCG alignment angle $\phi_{\rm bcg}$ is plotted with the x-axis divided into 9 bins, and the y-axis showing the normalized galaxy count per bin. The left, middle, and right panels show the statistical results for the total, inner, and outer samples, respectively. Red squares and blue triangles connected by dotted lines represent the statistical results of BA angles for red disk galaxies (RDGs) and blue disk galaxies (BDGs), respectively; while red squares and blue triangles connected by solid lines represent the statistical results for red elliptical galaxies (REGs) and blue elliptical galaxies (BEGs). The mean values ($\langle \phi_{\rm bcg} \rangle$) are shown in the lower left corner for disk galaxies and in the lower right corner for elliptical galaxies.  
    \label{fig:4.4}}
\end{figure*}  

(\romannumeral 6) We analyze the dependence of the BA signal on the fraction of QGs among the member galaxies of each cluster, with the results shown in Figure~\hyperref[fig:4.3]{\ref*{fig:4.3}(f)}. 
Overall, in clusters with a higher fraction of QGs, the alignment signal is more strongly enhanced for QGs, while SFGs also exhibit a clear increasing trend.   
Generally speaking, clusters with longer relaxation times tend to have a higher fraction of QGs (\citealt{2024MNRAS.529.3651O}). 
Results from simulations further reveal that 85\% of cluster QGs are ancient infallers, which is a consequence of cluster relaxation (\citealt{2023MNRAS.518.2398H}). 
Meanwhile, the feedback mechanism from central active galactic nuclei (AGN) associated with the BCG can influence the distribution of hot gas within the cluster (\citealt{2009Natur.460..213C}). 
This energy feedback reduces the density of the intracluster medium along the minor axis, weakening the efficiency of ram-pressure stripping (\citealt{2021Natur.594..187M}). 
As a result, a higher fraction of quiescent satellite galaxies is preferentially located along the major axis, thereby strengthening the BA signal.

\subsection{ Color-Morphology Decoupling in the BCG Alignment Signal }
\label{subsec:4.4}

As an increasing number of studies point out the decoupling between galaxy morphology evolution and color evolution, we seek to investigate the dependence of the BA signal on galaxy morphology. 
By measuring the BA signals of two special types of galaxies, namely red disk galaxies (RDGs) and blue elliptical galaxies (BEGs), it is verified that the effect of galaxy morphology differences on the alignment signal is limited.   
We further construct corresponding control samples of blue disk galaxies (BDGs) for RDGs and red elliptical galaxies (REGs) for BEGs. 
As shown in Figures~\ref{fig:4.4}, for red galaxies (i.e., QGs), the BA angle distributions of both disk (dotted line) and elliptical (solid line) galaxies are significantly concentrated below $45^\circ$; for blue galaxies (i.e., SFGs), however, neither galaxy type exhibits a pronounced BA signal.   

The analysis reveals that the BA signal is most significant in red-colored galaxies, regardless of whether they are morphologically classified as disk or elliptical galaxies. 
Furthermore, by analyzing two special galaxy populations, we confirm that this dependence on satellite galaxy color is not governed by that on satellite morphology, consistent with the results shown in Figure~\ref{fig:4.3}, where red galaxies exhibit a stronger signal than blue galaxies across all parameters.

\section{Summary}
\label{sec:5}
Analyzing the spatial distribution of member galaxies in clusters using observational data is an important tool for studying cluster dynamics and provides crucial observational constraints for galaxy formation models. 
In this paper, we use DESI-based photometric catalog data to systematically analyze the spatial distribution and alignment characteristics of SFGs and QGs in clusters. 
Leveraging the large-sample advantage of photometric data, we simultaneously employ two strict criteria, sSFR and color ($g-z$), to construct a sufficiently large analysis sample for an in-depth investigation of the alignment signals between these two types of samples. 
The main conclusions of this paper are summarized as follows: 
\begin{enumerate}
\item Based on the SFG/QG classification, we confirm that the BA signal strongly depends on satellite galaxy color, and this dependence is morphology-independent. Overall, QGs exhibit a strong BA signal, whereas SFGs show it mainly in cluster outskirts, with no significant signal in the inner regions. 
\item The dependence of the BA signal on various galactic and cluster properties reveals that it is significantly influenced by cluster richness and the QG fraction. This implies that the BA signal of satellite galaxies is more pronounced in more mature clusters with longer relaxation times. Analysis of satellite galaxy absolute magnitudes and axis ratios shows that QGs have a stronger signal when they are brighter and have larger axis ratios, while SFGs show no significant dependence. 
\item Examining the BA signal as a function of satellite galaxy radial distance reveals opposite trends for SFGs and QGs: the SFG signal is stronger in the outskirts, while the QG signal is stronger in the inner regions of the cluster. This reflects how the transport of galaxies along cosmic filaments drives cluster evolution: early-stage SFGs fall into clusters along anisotropic filaments, and as they fall toward the cluster center, their spatial orientations and distribution properties are reshaped by gravitational and tidal forces, which weakens the BA signal. QGs, which are early inhabitants within clusters and have long been exposed to high-density environments, experience stronger tidal forces and a longer virialization process, and thus show more pronounced alignment.  
\item When examining the SA signal, we find that the orientations of satellite galaxies of both SFGs and QGs in the outer regions of clusters tend to be isotropic, while in the inner regions, SFGs and QGs exhibit opposite trends. We interpret the observed anti-correlation in the SA signal in the cluster inner regions as being collectively explained by these combined factors: gravitational lensing, gravity, and tidal forces twisting. 
\end{enumerate}

In summary, QGs, as early cluster members, exhibit stable spatial orientations and distributions with significant anisotropy. 
In contrast, SFGs, as the new-generation cluster members, have their spatial orientations and distributions significantly modulated by both cosmic filaments and the intracluster environment, resulting in pronounced differences between cluster inner and outer regions.

\begin{acknowledgements} 

We thank the anonymous referee for the thorough, thoughtful review and professional suggestions, which have significantly improved the clarity and depth of this manuscript. 

This research is supported by the National Key R\&D Program of China (No. 2023YFA1607804) and the National Natural Science Foundation of China (NSFC; Nos. 12273013, 12120101003, 12373010, 12233008, and 12503013). 
H.Z. acknowledges the support from the National R\&D Program of China (Nos. 2022YFA1602902, 2023YFA1608100, and 2025YFE0202300) and the Strategic Priority Research Program of the Chinese Academy of Sciences (Grant Nos. XDB0550100 and XDB0550000), the China Manned Space Project (No. CMS-CSST-2025-A06), and the National Astronomical Observatories, Chinese Academy of Sciences (Nos. E5ZQ7801, E5ZB7801, and E4TG2001). 
Z.C. acknowledges the support from the Postgraduate Research \& Practice Innovation Program of Jiangsu Province (No. KYCX25\_1940).

The DESI Legacy Imaging Surveys consist of three individual and complementary projects: the Dark Energy Camera Legacy Survey (DECaLS), the Beijing-Arizona Sky Survey (BASS), and the Mayall z-band Legacy Survey (MzLS). DECaLS, BASS and MzLS together include data obtained, respectively, at the Blanco telescope, Cerro Tololo Inter-American Observatory, NSF’s NOIRLab; the Bok telescope, Steward Observatory, University of Arizona; and the Mayall telescope, Kitt Peak National Observatory, NOIRLab. NOIRLab is operated by the Association of Universities for Research in Astronomy (AURA) under a cooperative agreement with the National Science Foundation. Pipeline processing and analyzes of the data were supported by NOIRLab and the Lawrence Berkeley National Laboratory (LBNL). Legacy Surveys also uses data products from the Near-Earth Object Wide-field Infrared Survey Explorer (NEOWISE), a project of the Jet Propulsion Laboratory/California Institute of Technology, funded by the National Aeronautics and Space Administration. Legacy Surveys was supported by: the Director, Office of Science, Office of High Energy Physics of the U.S. Department of Energy; the National Energy Research Scientific Computing Center, a DOE Office of Science User Facility; the U.S. National Science Foundation, Division of Astronomical Sciences; the National Astronomical Observatories of China, the Chinese Academy of Sciences and the Chinese National Natural Science Foundation. LBNL is managed by the Regents of the University of California under contract to the U.S. Department of Energy.
   
\end{acknowledgements}


\bibliography{bibtex}

@ARTICLE{1969ArA.....5..305H,
       author = {{Holmberg}, E.},
        title = "{A study of physical groups of galaxies}",
      journal = {Arkiv for Astronomi},
         year = 1969,
        month = jan,
       volume = {5},
        pages = {305-343},
       adsurl = {https://ui.adsabs.harvard.edu/abs/1969ArA.....5..305H}
}

@ARTICLE{1971ApJ...165L..27G,
       author = {{Giacconi}, R. and {Kellogg}, E. and {Gorenstein}, P. and {Gursky}, H. and {Tananbaum}, H.},
        title = "{An X-Ray Scan of the Galactic Plane from UHURU}",
      journal = {\apjl},
         year = 1971,
        month = apr,
       volume = {165},
        pages = {L27},
          doi = {10.1086/180711},
       adsurl = {https://ui.adsabs.harvard.edu/abs/1971ApJ...165L..27G}
}

@ARTICLE{1972CoASP...4..173S,
       author = {{Sunyaev}, R.~A. and {Zeldovich}, Ya. B.},
        title = "{The Observations of Relic Radiation as a Test of the Nature of X-Ray Radiation from the Clusters of Galaxies}",
      journal = {Comments on Astrophysics and Space Physics},
         year = 1972,
        month = nov,
       volume = {4},
        pages = {173},
       adsurl = {https://ui.adsabs.harvard.edu/abs/1972CoASP...4..173S}
}

@ARTICLE{1984ApJ...286..186F,
       author = {{Fabricant}, D. and {Rybicki}, G. and {Gorenstein}, P.},
        title = "{X-ray measurements of the nonspherical mass distribution in the cluster of galaxies A 2256.}",
      journal = {\apj},
         year = 1984,
        month = nov,
       volume = {286},
        pages = {186-195},
          doi = {10.1086/162586},
       adsurl = {https://ui.adsabs.harvard.edu/abs/1984ApJ...286..186F}
}

@ARTICLE{1994MNRAS.270..390C,
       author = {{Ciotti}, L. and {Dutta}, S.~N.},
        title = "{Alignment and morphology of elliptical galaxies : the influence of the cluster tidal field.}",
      journal = {\mnras},
         year = 1994,
        month = sep,
       volume = {270},
        pages = {390-400},
          doi = {10.1093/mnras/270.2.390},
archivePrefix = {arXiv},
       eprint = {astro-ph/9404059},
 primaryClass = {astro-ph},
       adsurl = {https://ui.adsabs.harvard.edu/abs/1994MNRAS.270..390C}
}

@ARTICLE{2006MNRAS.369.1293Y,
       author = {{Yang}, Xiaohu and {van den Bosch}, Frank C. and {Mo}, H.~J. and {Mao}, Shude and {Kang}, Xi and {Weinmann}, Simone M. and {Guo}, Yicheng and {Jing}, Y.~P.},
        title = "{The alignment between the distribution of satellites and the orientation of their central galaxy}",
      journal = {\mnras},
         year = 2006,
        month = jul,
       volume = {369},
       number = {3},
        pages = {1293-1302},
          doi = {10.1111/j.1365-2966.2006.10373.x},
archivePrefix = {arXiv},
       eprint = {astro-ph/0601040},
 primaryClass = {astro-ph},
       adsurl = {https://ui.adsabs.harvard.edu/abs/2006MNRAS.369.1293Y}
}

@ARTICLE{2008ApJ...675..146F,
       author = {{Faltenbacher}, A. and {Jing}, Y.~P. and {Li}, Cheng and {Mao}, Shude and {Mo}, H.~J. and {Pasquali}, Anna and {van den Bosch}, Frank C.},
        title = "{Spatial and Kinematic Alignments between Central and Satellite Halos}",
      journal = {\apj},
         year = 2008,
        month = mar,
       volume = {675},
       number = {1},
        pages = {146-155},
          doi = {10.1086/525243},
archivePrefix = {arXiv},
       eprint = {0706.0262},
 primaryClass = {astro-ph},
       adsurl = {https://ui.adsabs.harvard.edu/abs/2008ApJ...675..146F}
}

@ARTICLE{2010PASJ...62..811O,
       author = {{Okabe}, Nobuhiro and {Takada}, Masahiro and {Umetsu}, Keiichi and {Futamase}, Toshifumi and {Smith}, Graham P.},
        title = "{LoCuSS: Subaru Weak Lensing Study of 30 Galaxy Clusters}",
      journal = {\pasj},
         year = 2010,
        month = jun,
       volume = {62},
        pages = {811},
          doi = {10.1093/pasj/62.3.811},
archivePrefix = {arXiv},
       eprint = {0903.1103},
 primaryClass = {astro-ph.CO},
       adsurl = {https://ui.adsabs.harvard.edu/abs/2010PASJ...62..811O}
}

@ARTICLE{2010MNRAS.405.2023N,
       author = {{Niederste-Ostholt}, Martin and {Strauss}, Michael A. and {Dong}, Feng and {Koester}, Benjamin P. and {McKay}, Timothy A.},
        title = "{Alignment of brightest cluster galaxies with their host clusters}",
      journal = {\mnras},
         year = 2010,
        month = jul,
       volume = {405},
       number = {3},
        pages = {2023-2036},
          doi = {10.1111/j.1365-2966.2010.16597.x},
archivePrefix = {arXiv},
       eprint = {1003.0322},
 primaryClass = {astro-ph.CO},
       adsurl = {https://ui.adsabs.harvard.edu/abs/2010MNRAS.405.2023N}
}

@ARTICLE{2015MNRAS.450.2195S,
       author = {{Singh}, Sukhdeep and {Mandelbaum}, Rachel and {More}, Surhud},
        title = "{Intrinsic alignments of SDSS-III BOSS LOWZ sample galaxies}",
      journal = {\mnras},
         year = 2015,
        month = jun,
       volume = {450},
       number = {2},
        pages = {2195-2216},
          doi = {10.1093/mnras/stv778},
archivePrefix = {arXiv},
       eprint = {1411.1755},
 primaryClass = {astro-ph.CO},
       adsurl = {https://ui.adsabs.harvard.edu/abs/2015MNRAS.450.2195S}
}

@ARTICLE{2016MNRAS.463..222H,
       author = {{Huang}, Hung-Jin and {Mandelbaum}, Rachel and {Freeman}, Peter E. and {Chen}, Yen-Chi and {Rozo}, Eduardo and {Rykoff}, Eli and {Baxter}, Eric J.},
        title = "{Intrinsic alignments in redMaPPer clusters - I. Central galaxy alignments and angular segregation of satellites}",
      journal = {\mnras},
         year = 2016,
        month = nov,
       volume = {463},
       number = {1},
        pages = {222-244},
          doi = {10.1093/mnras/stw1982},
archivePrefix = {arXiv},
       eprint = {1605.01065},
 primaryClass = {astro-ph.GA},
       adsurl = {https://ui.adsabs.harvard.edu/abs/2016MNRAS.463..222H}
}

@ARTICLE{2018MNRAS.474.4772H,
       author = {{Huang}, Hung-Jin and {Mandelbaum}, Rachel and {Freeman}, Peter E. and {Chen}, Yen-Chi and {Rozo}, Eduardo and {Rykoff}, Eli},
        title = "{Intrinsic alignment in redMaPPer clusters - II. Radial alignment of satellites towards cluster centres}",
      journal = {\mnras},
         year = 2018,
        month = mar,
       volume = {474},
       number = {4},
        pages = {4772-4794},
          doi = {10.1093/mnras/stx2995},
archivePrefix = {arXiv},
       eprint = {1704.06273},
 primaryClass = {astro-ph.GA},
       adsurl = {https://ui.adsabs.harvard.edu/abs/2018MNRAS.474.4772H}
}

@ARTICLE{2022MNRAS.516.3159Y,
       author = {{Yuan}, Z.~S. and {Wen}, Z.~L.},
        title = "{The alignment between brightest cluster galaxies and host clusters}",
      journal = {\mnras},
         year = 2022,
        month = nov,
       volume = {516},
       number = {3},
        pages = {3159-3166},
          doi = {10.1093/mnras/stac2492},
archivePrefix = {arXiv},
       eprint = {2209.00204},
 primaryClass = {astro-ph.GA},
       adsurl = {https://ui.adsabs.harvard.edu/abs/2022MNRAS.516.3159Y}
}

@ARTICLE{2021ApJS..253...56Z,
       author = {{Zou}, Hu and {Gao}, Jinghua and {Xu}, Xin and {Zhou}, Xu and {Ma}, Jun and {Zhou}, Zhimin and {Zhang}, Tianmeng and {Nie}, Jundan and {Wang}, Jiali and {Xue}, Suijian},
        title = "{Galaxy Clusters from the DESI Legacy Imaging Surveys. I. Cluster Detection}",
      journal = {\apjs},
         year = 2021,
        month = apr,
       volume = {253},
       number = {2},
          eid = {56},
        pages = {56},
          doi = {10.3847/1538-4365/abe5b0},
archivePrefix = {arXiv},
       eprint = {2101.12340},
 primaryClass = {astro-ph.GA},
       adsurl = {https://ui.adsabs.harvard.edu/abs/2021ApJS..253...56Z}
}

@ARTICLE{2024ApJ...961..253C,
       author = {{Chen}, Zhaobin and {Gu}, Yizhou and {Zou}, Hu and {Yuan}, Qirong},
        title = "{Galaxy Clusters from the DESI Legacy Imaging Surveys. II. Environmental Effects on the Size{\textendash}Mass Relation}",
      journal = {\apj},
         year = 2024,
        month = feb,
       volume = {961},
       number = {2},
          eid = {253},
        pages = {253},
          doi = {10.3847/1538-4357/ad15fd},
archivePrefix = {arXiv},
       eprint = {2312.17075},
 primaryClass = {astro-ph.GA},
       adsurl = {https://ui.adsabs.harvard.edu/abs/2024ApJ...961..253C}
}

@ARTICLE{2019AJ....157..168D,
       author = {{Dey}, Arjun and {Schlegel}, David J. and {Lang}, Dustin and {Blum}, Robert and {Burleigh}, Kaylan and {Fan}, Xiaohui and {Findlay}, Joseph R. and {Finkbeiner}, Doug and {Herrera}, David and {Juneau}, St{\'e}phanie and {Landriau}, Martin and {Levi}, Michael and {McGreer}, Ian and {Meisner}, Aaron and {Myers}, Adam D. and {Moustakas}, John and {Nugent}, Peter and {Patej}, Anna and {Schlafly}, Edward F. and {Walker}, Alistair R. and {Valdes}, Francisco and {Weaver}, Benjamin A. and {Y{\`e}che}, Christophe and {Zou}, Hu and {Zhou}, Xu and {Abareshi}, Behzad and {Abbott}, T.~M.~C. and {Abolfathi}, Bela and {Aguilera}, C. and {Alam}, Shadab and {Allen}, Lori and {Alvarez}, A. and {Annis}, James and {Ansarinejad}, Behzad and {Aubert}, Marie and {Beechert}, Jacqueline and {Bell}, Eric F. and {BenZvi}, Segev Y. and {Beutler}, Florian and {Bielby}, Richard M. and {Bolton}, Adam S. and {Brice{\~n}o}, C{\'e}sar and {Buckley-Geer}, Elizabeth J. and {Butler}, Karen and {Calamida}, Annalisa and {Carlberg}, Raymond G. and {Carter}, Paul and {Casas}, Ricard and {Castander}, Francisco J. and {Choi}, Yumi and {Comparat}, Johan and {Cukanovaite}, Elena and {Delubac}, Timoth{\'e}e and {DeVries}, Kaitlin and {Dey}, Sharmila and {Dhungana}, Govinda and {Dickinson}, Mark and {Ding}, Zhejie and {Donaldson}, John B. and {Duan}, Yutong and {Duckworth}, Christopher J. and {Eftekharzadeh}, Sarah and {Eisenstein}, Daniel J. and {Etourneau}, Thomas and {Fagrelius}, Parker A. and {Farihi}, Jay and {Fitzpatrick}, Mike and {Font-Ribera}, Andreu and {Fulmer}, Leah and {G{\"a}nsicke}, Boris T. and {Gaztanaga}, Enrique and {George}, Koshy and {Gerdes}, David W. and {Gontcho}, Satya Gontcho A. and {Gorgoni}, Claudio and {Green}, Gregory and {Guy}, Julien and {Harmer}, Diane and {Hernandez}, M. and {Honscheid}, Klaus and {Huang}, Lijuan Wendy and {James}, David J. and {Jannuzi}, Buell T. and {Jiang}, Linhua and {Joyce}, Richard and {Karcher}, Armin and {Karkar}, Sonia and {Kehoe}, Robert and {Kneib}, Jean-Paul and {Kueter-Young}, Andrea and {Lan}, Ting-Wen and {Lauer}, Tod R. and {Le Guillou}, Laurent and {Le Van Suu}, Auguste and {Lee}, Jae Hyeon and {Lesser}, Michael and {Perreault Levasseur}, Laurence and {Li}, Ting S. and {Mann}, Justin L. and {Marshall}, Robert and {Mart{\'\i}nez-V{\'a}zquez}, C.~E. and {Martini}, Paul and {du Mas des Bourboux}, H{\'e}lion and {McManus}, Sean and {Meier}, Tobias Gabriel and {M{\'e}nard}, Brice and {Metcalfe}, Nigel and {Mu{\~n}oz-Guti{\'e}rrez}, Andrea and {Najita}, Joan and {Napier}, Kevin and {Narayan}, Gautham and {Newman}, Jeffrey A. and {Nie}, Jundan and {Nord}, Brian and {Norman}, Dara J. and {Olsen}, Knut A.~G. and {Paat}, Anthony and {Palanque-Delabrouille}, Nathalie and {Peng}, Xiyan and {Poppett}, Claire L. and {Poremba}, Megan R. and {Prakash}, Abhishek and {Rabinowitz}, David and {Raichoor}, Anand and {Rezaie}, Mehdi and {Robertson}, A.~N. and {Roe}, Natalie A. and {Ross}, Ashley J. and {Ross}, Nicholas P. and {Rudnick}, Gregory and {Safonova}, Sasha and {Saha}, Abhijit and {S{\'a}nchez}, F. Javier and {Savary}, Elodie and {Schweiker}, Heidi and {Scott}, Adam and {Seo}, Hee-Jong and {Shan}, Huanyuan and {Silva}, David R. and {Slepian}, Zachary and {Soto}, Christian and {Sprayberry}, David and {Staten}, Ryan and {Stillman}, Coley M. and {Stupak}, Robert J. and {Summers}, David L. and {Sien Tie}, Suk and {Tirado}, H. and {Vargas-Maga{\~n}a}, Mariana and {Vivas}, A. Katherina and {Wechsler}, Risa H. and {Williams}, Doug and {Yang}, Jinyi and {Yang}, Qian and {Yapici}, Tolga and {Zaritsky}, Dennis and {Zenteno}, A. and {Zhang}, Kai and {Zhang}, Tianmeng and {Zhou}, Rongpu and {Zhou}, Zhimin},
        title = "{Overview of the DESI Legacy Imaging Surveys}",
      journal = {\aj},
         year = 2019,
        month = may,
       volume = {157},
       number = {5},
          eid = {168},
        pages = {168},
          doi = {10.3847/1538-3881/ab089d},
archivePrefix = {arXiv},
       eprint = {1804.08657},
 primaryClass = {astro-ph.IM},
       adsurl = {https://ui.adsabs.harvard.edu/abs/2019AJ....157..168D}
}

@ARTICLE{2016MNRAS.460.1270D,
       author = {{Dark Energy Survey Collaboration} and {Abbott}, T. and {Abdalla}, F.~B. and {Aleksi{\'c}}, J. and {Allam}, S. and {Amara}, A. and {Bacon}, D. and {Balbinot}, E. and {Banerji}, M. and {Bechtol}, K. and {Benoit-L{\'e}vy}, A. and {Bernstein}, G.~M. and {Bertin}, E. and {Blazek}, J. and {Bonnett}, C. and {Bridle}, S. and {Brooks}, D. and {Brunner}, R.~J. and {Buckley-Geer}, E. and {Burke}, D.~L. and {Caminha}, G.~B. and {Capozzi}, D. and {Carlsen}, J. and {Carnero-Rosell}, A. and {Carollo}, M. and {Carrasco-Kind}, M. and {Carretero}, J. and {Castander}, F.~J. and {Clerkin}, L. and {Collett}, T. and {Conselice}, C. and {Crocce}, M. and {Cunha}, C.~E. and {D'Andrea}, C.~B. and {da Costa}, L.~N. and {Davis}, T.~M. and {Desai}, S. and {Diehl}, H.~T. and {Dietrich}, J.~P. and {Dodelson}, S. and {Doel}, P. and {Drlica-Wagner}, A. and {Estrada}, J. and {Etherington}, J. and {Evrard}, A.~E. and {Fabbri}, J. and {Finley}, D.~A. and {Flaugher}, B. and {Foley}, R.~J. and {Fosalba}, P. and {Frieman}, J. and {Garc{\'\i}a-Bellido}, J. and {Gaztanaga}, E. and {Gerdes}, D.~W. and {Giannantonio}, T. and {Goldstein}, D.~A. and {Gruen}, D. and {Gruendl}, R.~A. and {Guarnieri}, P. and {Gutierrez}, G. and {Hartley}, W. and {Honscheid}, K. and {Jain}, B. and {James}, D.~J. and {Jeltema}, T. and {Jouvel}, S. and {Kessler}, R. and {King}, A. and {Kirk}, D. and {Kron}, R. and {Kuehn}, K. and {Kuropatkin}, N. and {Lahav}, O. and {Li}, T.~S. and {Lima}, M. and {Lin}, H. and {Maia}, M.~A.~G. and {Makler}, M. and {Manera}, M. and {Maraston}, C. and {Marshall}, J.~L. and {Martini}, P. and {McMahon}, R.~G. and {Melchior}, P. and {Merson}, A. and {Miller}, C.~J. and {Miquel}, R. and {Mohr}, J.~J. and {Morice-Atkinson}, X. and {Naidoo}, K. and {Neilsen}, E. and {Nichol}, R.~C. and {Nord}, B. and {Ogando}, R. and {Ostrovski}, F. and {Palmese}, A. and {Papadopoulos}, A. and {Peiris}, H.~V. and {Peoples}, J. and {Percival}, W.~J. and {Plazas}, A.~A. and {Reed}, S.~L. and {Refregier}, A. and {Romer}, A.~K. and {Roodman}, A. and {Ross}, A. and {Rozo}, E. and {Rykoff}, E.~S. and {Sadeh}, I. and {Sako}, M. and {S{\'a}nchez}, C. and {Sanchez}, E. and {Santiago}, B. and {Scarpine}, V. and {Schubnell}, M. and {Sevilla-Noarbe}, I. and {Sheldon}, E. and {Smith}, M. and {Smith}, R.~C. and {Soares-Santos}, M. and {Sobreira}, F. and {Soumagnac}, M. and {Suchyta}, E. and {Sullivan}, M. and {Swanson}, M. and {Tarle}, G. and {Thaler}, J. and {Thomas}, D. and {Thomas}, R.~C. and {Tucker}, D. and {Vieira}, J.~D. and {Vikram}, V. and {Walker}, A.~R. and {Wechsler}, R.~H. and {Weller}, J. and {Wester}, W. and {Whiteway}, L. and {Wilcox}, H. and {Yanny}, B. and {Zhang}, Y. and {Zuntz}, J.},
        title = "{The Dark Energy Survey: more than dark energy - an overview}",
      journal = {\mnras},
         year = 2016,
        month = aug,
       volume = {460},
       number = {2},
        pages = {1270-1299},
          doi = {10.1093/mnras/stw641},
archivePrefix = {arXiv},
       eprint = {1601.00329},
 primaryClass = {astro-ph.CO},
       adsurl = {https://ui.adsabs.harvard.edu/abs/2016MNRAS.460.1270D}
}

@ARTICLE{2016arXiv161100036D,
       author = {{DESI Collaboration} and {Aghamousa}, Amir and {Aguilar}, Jessica and {Ahlen}, Steve and {Alam}, Shadab and {Allen}, Lori E. and {Allende Prieto}, Carlos and {Annis}, James and {Bailey}, Stephen and {Balland}, Christophe and {Ballester}, Otger and {Baltay}, Charles and {Beaufore}, Lucas and {Bebek}, Chris and {Beers}, Timothy C. and {Bell}, Eric F. and {Bernal}, Jos{\'e} Luis and {Besuner}, Robert and {Beutler}, Florian and {Blake}, Chris and {Bleuler}, Hannes and {Blomqvist}, Michael and {Blum}, Robert and {Bolton}, Adam S. and {Briceno}, Cesar and {Brooks}, David and {Brownstein}, Joel R. and {Buckley-Geer}, Elizabeth and {Burden}, Angela and {Burtin}, Etienne and {Busca}, Nicolas G. and {Cahn}, Robert N. and {Cai}, Yan-Chuan and {Cardiel-Sas}, Laia and {Carlberg}, Raymond G. and {Carton}, Pierre-Henri and {Casas}, Ricard and {Castander}, Francisco J. and {Cervantes-Cota}, Jorge L. and {Claybaugh}, Todd M. and {Close}, Madeline and {Coker}, Carl T. and {Cole}, Shaun and {Comparat}, Johan and {Cooper}, Andrew P. and {Cousinou}, M. -C. and {Crocce}, Martin and {Cuby}, Jean-Gabriel and {Cunningham}, Daniel P. and {Davis}, Tamara M. and {Dawson}, Kyle S. and {de la Macorra}, Axel and {De Vicente}, Juan and {Delubac}, Timoth{\'e}e and {Derwent}, Mark and {Dey}, Arjun and {Dhungana}, Govinda and {Ding}, Zhejie and {Doel}, Peter and {Duan}, Yutong T. and {Ealet}, Anne and {Edelstein}, Jerry and {Eftekharzadeh}, Sarah and {Eisenstein}, Daniel J. and {Elliott}, Ann and {Escoffier}, St{\'e}phanie and {Evatt}, Matthew and {Fagrelius}, Parker and {Fan}, Xiaohui and {Fanning}, Kevin and {Farahi}, Arya and {Farihi}, Jay and {Favole}, Ginevra and {Feng}, Yu and {Fernandez}, Enrique and {Findlay}, Joseph R. and {Finkbeiner}, Douglas P. and {Fitzpatrick}, Michael J. and {Flaugher}, Brenna and {Flender}, Samuel and {Font-Ribera}, Andreu and {Forero-Romero}, Jaime E. and {Fosalba}, Pablo and {Frenk}, Carlos S. and {Fumagalli}, Michele and {Gaensicke}, Boris T. and {Gallo}, Giuseppe and {Garcia-Bellido}, Juan and {Gaztanaga}, Enrique and {Pietro Gentile Fusillo}, Nicola and {Gerard}, Terry and {Gershkovich}, Irena and {Giannantonio}, Tommaso and {Gillet}, Denis and {Gonzalez-de-Rivera}, Guillermo and {Gonzalez-Perez}, Violeta and {Gott}, Shelby and {Graur}, Or and {Gutierrez}, Gaston and {Guy}, Julien and {Habib}, Salman and {Heetderks}, Henry and {Heetderks}, Ian and {Heitmann}, Katrin and {Hellwing}, Wojciech A. and {Herrera}, David A. and {Ho}, Shirley and {Holland}, Stephen and {Honscheid}, Klaus and {Huff}, Eric and {Hutchinson}, Timothy A. and {Huterer}, Dragan and {Hwang}, Ho Seong and {Illa Laguna}, Joseph Maria and {Ishikawa}, Yuzo and {Jacobs}, Dianna and {Jeffrey}, Niall and {Jelinsky}, Patrick and {Jennings}, Elise and {Jiang}, Linhua and {Jimenez}, Jorge and {Johnson}, Jennifer and {Joyce}, Richard and {Jullo}, Eric and {Juneau}, St{\'e}phanie and {Kama}, Sami and {Karcher}, Armin and {Karkar}, Sonia and {Kehoe}, Robert and {Kennamer}, Noble and {Kent}, Stephen and {Kilbinger}, Martin and {Kim}, Alex G. and {Kirkby}, David and {Kisner}, Theodore and {Kitanidis}, Ellie and {Kneib}, Jean-Paul and {Koposov}, Sergey and {Kovacs}, Eve and {Koyama}, Kazuya and {Kremin}, Anthony and {Kron}, Richard and {Kronig}, Luzius and {Kueter-Young}, Andrea and {Lacey}, Cedric G. and {Lafever}, Robin and {Lahav}, Ofer and {Lambert}, Andrew and {Lampton}, Michael and {Landriau}, Martin and {Lang}, Dustin and {Lauer}, Tod R. and {Le Goff}, Jean-Marc and {Le Guillou}, Laurent and {Le Van Suu}, Auguste and {Lee}, Jae Hyeon and {Lee}, Su-Jeong and {Leitner}, Daniela and {Lesser}, Michael and {Levi}, Michael E. and {L'Huillier}, Benjamin and {Li}, Baojiu and {Liang}, Ming and {Lin}, Huan and {Linder}, Eric and {Loebman}, Sarah R. and {Luki{\'c}}, Zarija and {Ma}, Jun and {MacCrann}, Niall and {Magneville}, Christophe and {Makarem}, Laleh and {Manera}, Marc and {Manser}, Christopher J. and {Marshall}, Robert and {Martini}, Paul and {Massey}, Richard and {Matheson}, Thomas and {McCauley}, Jeremy and {McDonald}, Patrick and {McGreer}, Ian D. and {Meisner}, Aaron and {Metcalfe}, Nigel and {Miller}, Timothy N. and {Miquel}, Ramon and {Moustakas}, John and {Myers}, Adam and {Naik}, Milind and {Newman}, Jeffrey A. and {Nichol}, Robert C. and {Nicola}, Andrina and {Nicolati da Costa}, Luiz and {Nie}, Jundan and {Niz}, Gustavo and {Norberg}, Peder and {Nord}, Brian and {Norman}, Dara and {Nugent}, Peter and {O'Brien}, Thomas and {Oh}, Minji and {Olsen}, Knut A.~G.},
        title = "{The DESI Experiment Part I: Science,Targeting, and Survey Design}",
      journal = {arXiv e-prints},
         year = 2016,
        month = oct,
          eid = {arXiv:1611.00036},
        pages = {arXiv:1611.00036},
          doi = {10.48550/arXiv.1611.00036},
archivePrefix = {arXiv},
       eprint = {1611.00036},
 primaryClass = {astro-ph.IM},
       adsurl = {https://ui.adsabs.harvard.edu/abs/2016arXiv161100036D}
}

@ARTICLE{2010AJ....140.1868W,
       author = {{Wright}, Edward L. and {Eisenhardt}, Peter R.~M. and {Mainzer}, Amy K. and {Ressler}, Michael E. and {Cutri}, Roc M. and {Jarrett}, Thomas and {Kirkpatrick}, J. Davy and {Padgett}, Deborah and {McMillan}, Robert S. and {Skrutskie}, Michael and {Stanford}, S.~A. and {Cohen}, Martin and {Walker}, Russell G. and {Mather}, John C. and {Leisawitz}, David and {Gautier}, III, Thomas N. and {McLean}, Ian and {Benford}, Dominic and {Lonsdale}, Carol J. and {Blain}, Andrew and {Mendez}, Bryan and {Irace}, William R. and {Duval}, Valerie and {Liu}, Fengchuan and {Royer}, Don and {Heinrichsen}, Ingolf and {Howard}, Joan and {Shannon}, Mark and {Kendall}, Martha and {Walsh}, Amy L. and {Larsen}, Mark and {Cardon}, Joel G. and {Schick}, Scott and {Schwalm}, Mark and {Abid}, Mohamed and {Fabinsky}, Beth and {Naes}, Larry and {Tsai}, Chao-Wei},
        title = "{The Wide-field Infrared Survey Explorer (WISE): Mission Description and Initial On-orbit Performance}",
      journal = {\aj},
         year = 2010,
        month = dec,
       volume = {140},
       number = {6},
        pages = {1868-1881},
          doi = {10.1088/0004-6256/140/6/1868},
archivePrefix = {arXiv},
       eprint = {1008.0031},
 primaryClass = {astro-ph.IM},
       adsurl = {https://ui.adsabs.harvard.edu/abs/2010AJ....140.1868W}
}

@software{2016ascl.soft04008L,
       author = {{Lang}, Dustin and {Hogg}, David W. and {Mykytyn}, David},
        title = "{The Tractor: Probabilistic astronomical source detection and measurement}",
 howpublished = {Astrophysics Source Code Library, record ascl:1604.008},
         year = 2016,
        month = apr,
          eid = {ascl:1604.008},
       adsurl = {https://ui.adsabs.harvard.edu/abs/2016ascl.soft04008L}
}

@INPROCEEDINGS{2011ASPC..442..435B,
       author = {{Bertin}, E.},
        title = "{Automated Morphometry with SExtractor and PSFEx}",
    booktitle = {Astronomical Data Analysis Software and Systems XX},
         year = 2011,
       editor = {{Evans}, I.~N. and {Accomazzi}, A. and {Mink}, D.~J. and {Rots}, A.~H.},
       series = {Astronomical Society of the Pacific Conference Series},
       volume = {442},
        month = jul,
        pages = {435},
       adsurl = {https://ui.adsabs.harvard.edu/abs/2011ASPC..442..435B}
}

@ARTICLE{2020PASP..132b4101G,
       author = {{Gao}, Jinghua and {Zou}, Hu and {Zhou}, Xu and {Kong}, Xu},
        title = "{A Catalog of Galaxy Clusters Identified from SCUSS, SDSS, and UNWISE}",
      journal = {\pasp},
         year = 2020,
        month = feb,
       volume = {132},
       number = {1008},
          eid = {024101},
        pages = {024101},
          doi = {10.1088/1538-3873/ab6151},
archivePrefix = {arXiv},
       eprint = {1912.10909},
 primaryClass = {astro-ph.GA},
       adsurl = {https://ui.adsabs.harvard.edu/abs/2020PASP..132b4101G}
}

@ARTICLE{2019ApJS..242....8Z,
       author = {{Zou}, Hu and {Gao}, Jinghua and {Zhou}, Xu and {Kong}, Xu},
        title = "{Photometric Redshifts and Stellar Masses for Galaxies from the DESI Legacy Imaging Surveys}",
      journal = {\apjs},
         year = 2019,
        month = may,
       volume = {242},
       number = {1},
          eid = {8},
        pages = {8},
          doi = {10.3847/1538-4365/ab1847},
       adsurl = {https://ui.adsabs.harvard.edu/abs/2019ApJS..242....8Z}
}

@ARTICLE{2022RAA....22f5001Z,
       author = {{Zou}, Hu and {Sui}, Jipeng and {Xue}, Suijian and {Zhou}, Xu and {Ma}, Jun and {Zhou}, Zhimin and {Nie}, Jundan and {Zhang}, Tianmeng and {Feng}, Lu and {Shen}, Zhixia and {Wang}, Jiali},
        title = "{Photometric Redshifts and Galaxy Clusters for DES DR2, DESI DR9, and HSC-SSP PDR3 Data}",
      journal = {Research in Astronomy and Astrophysics},
         year = 2022,
        month = jun,
       volume = {22},
       number = {6},
          eid = {065001},
        pages = {065001},
          doi = {10.1088/1674-4527/ac6416},
archivePrefix = {arXiv},
       eprint = {2203.17035},
 primaryClass = {astro-ph.GA},
       adsurl = {https://ui.adsabs.harvard.edu/abs/2022RAA....22f5001Z}
}

@ARTICLE{2013ApJ...770L..12L,
       author = {{Li}, Cheng and {Jing}, Y.~P. and {Faltenbacher}, A. and {Wang}, Jie},
        title = "{The Detection of the Large-scale Alignment of Massive Galaxies at z \raisebox{-0.5ex}\textasciitilde 0.6}",
      journal = {\apjl},
         year = 2013,
        month = jun,
       volume = {770},
       number = {1},
          eid = {L12},
        pages = {L12},
          doi = {10.1088/2041-8205/770/1/L12},
archivePrefix = {arXiv},
       eprint = {1303.1965},
 primaryClass = {astro-ph.CO},
       adsurl = {https://ui.adsabs.harvard.edu/abs/2013ApJ...770L..12L}
}

@ARTICLE{2015SSRv..193....1J,
       author = {{Joachimi}, Benjamin and {Cacciato}, Marcello and {Kitching}, Thomas D. and {Leonard}, Adrienne and {Mandelbaum}, Rachel and {Sch{\"a}fer}, Bj{\"o}rn Malte and {Sif{\'o}n}, Crist{\'o}bal and {Hoekstra}, Henk and {Kiessling}, Alina and {Kirk}, Donnacha and et al.},
        title = "{Galaxy Alignments: An Overview}",
      journal = {\ssr},
         year = 2015,
        month = nov,
       volume = {193},
       number = {1-4},
        pages = {1-65},
          doi = {10.1007/s11214-015-0177-4},
archivePrefix = {arXiv},
       eprint = {1504.05456},
 primaryClass = {astro-ph.GA},
       adsurl = {https://ui.adsabs.harvard.edu/abs/2015SSRv..193....1J}
}

@ARTICLE{2015SSRv..193..139K,
       author = {{Kirk}, Donnacha and {Brown}, Michael L. and {Hoekstra}, Henk and {Joachimi}, Benjamin and {Kitching}, Thomas D. and {Mandelbaum}, Rachel and {Sif{\'o}n}, Crist{\'o}bal and {Cacciato}, Marcello and {Choi}, Ami and {Kiessling}, Alina and et al.},
        title = "{Galaxy Alignments: Observations and Impact on Cosmology}",
      journal = {\ssr},
         year = 2015,
        month = nov,
       volume = {193},
       number = {1-4},
        pages = {139-211},
          doi = {10.1007/s11214-015-0213-4},
archivePrefix = {arXiv},
       eprint = {1504.05465},
 primaryClass = {astro-ph.GA},
       adsurl = {https://ui.adsabs.harvard.edu/abs/2015SSRv..193..139K}
}

@ARTICLE{2013ApJ...779..160Z,
       author = {{Zhang}, Youcai and {Yang}, Xiaohu and {Wang}, Huiyuan and {Wang}, Lei and {Mo}, H.~J. and {van den Bosch}, Frank C.},
        title = "{Alignments of Galaxies within Cosmic Filaments from SDSS DR7}",
      journal = {\apj},
         year = 2013,
        month = dec,
       volume = {779},
       number = {2},
          eid = {160},
        pages = {160},
          doi = {10.1088/0004-637X/779/2/160},
archivePrefix = {arXiv},
       eprint = {1309.3847},
 primaryClass = {astro-ph.CO},
       adsurl = {https://ui.adsabs.harvard.edu/abs/2013ApJ...779..160Z}
}

@ARTICLE{2024ApJ...974...40L,
       author = {{Lan}, Yanyao and {Tang}, Lin and {Lin}, Weipeng and {Gong}, Junyu},
        title = "{Mock Observations: Three Different Types of Galaxy Alignment in TNG100 Simulations}",
      journal = {\apj},
         year = 2024,
        month = oct,
       volume = {974},
       number = {1},
          eid = {40},
        pages = {40},
          doi = {10.3847/1538-4357/ad6982},
archivePrefix = {arXiv},
       eprint = {2408.02230},
 primaryClass = {astro-ph.GA},
       adsurl = {https://ui.adsabs.harvard.edu/abs/2024ApJ...974...40L}
}

@ARTICLE{2025ApJ...983..122R,
       author = {{Rong}, Yu and {Wang}, Peng},
        title = "{Alignment of Blue/Green and Red Early-type Galaxies with Large-scale Filaments Reveals Distinct Evolutionary Pathways}",
      journal = {\apj},
         year = 2025,
        month = apr,
       volume = {983},
       number = {2},
          eid = {122},
        pages = {122},
          doi = {10.3847/1538-4357/adc4de},
archivePrefix = {arXiv},
       eprint = {2411.14885},
 primaryClass = {astro-ph.GA},
       adsurl = {https://ui.adsabs.harvard.edu/abs/2025ApJ...983..122R}
}

@ARTICLE{2023MNRAS.525.4685S,
       author = {{Smith}, Rory and {Hwang}, Ho Seong and {Kraljic}, Katarina and {Calder{\'o}n-Castillo}, Paula and {Jackson}, Thomas M. and {Pasquali}, Anna and {Shin}, Jihye and {Ko}, Jongwan and {Yoo}, Jaewon and {Kim}, Hyowon and et al.},
        title = "{BCG alignment with the locations of cluster members and the large-scale structure out to 10 R$_{200}$}",
      journal = {\mnras},
         year = 2023,
        month = nov,
       volume = {525},
       number = {3},
        pages = {4685-4699},
          doi = {10.1093/mnras/stad2535},
archivePrefix = {arXiv},
       eprint = {2308.08622},
 primaryClass = {astro-ph.GA},
       adsurl = {https://ui.adsabs.harvard.edu/abs/2023MNRAS.525.4685S}
}

@ARTICLE{2025A&A...700A.264D,
       author = {{Doubrawa}, L. and {Smith}, R. and {Mendes de Oliveira}, C. and {O'Mill}, A.~L. and {Nakazono}, L. and {Herpich}, F.~R. and {Gon{\c{c}}alves}, T.~S. and {Kanaan}, A. and {Ribeiro}, T. and {Schoenell}, W.},
        title = "{SCALE: Using S-PLUS photometric information to analyse the brightest cluster galaxy alignment with satellite galaxies}",
      journal = {\aap},
         year = 2025,
        month = aug,
       volume = {700},
          eid = {A264},
        pages = {A264},
          doi = {10.1051/0004-6361/202556043},
       adsurl = {https://ui.adsabs.harvard.edu/abs/2025A&A...700A.264D}
}

@ARTICLE{2009ApJ...694..214O,
       author = {{Okumura}, Teppei and {Jing}, Y.~P. and {Li}, Cheng},
        title = "{Intrinsic Ellipticity Correlation of SDSS Luminous Red Galaxies and Misalignment with Their Host Dark Matter Halos}",
      journal = {\apj},
         year = 2009,
        month = mar,
       volume = {694},
       number = {1},
        pages = {214-221},
          doi = {10.1088/0004-637X/694/1/214},
archivePrefix = {arXiv},
       eprint = {0809.3790},
 primaryClass = {astro-ph},
       adsurl = {https://ui.adsabs.harvard.edu/abs/2009ApJ...694..214O}
}

@ARTICLE{2015PhR...558....1T,
       author = {{Troxel}, M.~A. and {Ishak}, Mustapha},
        title = "{The intrinsic alignment of galaxies and its impact on weak gravitational lensing in an era of precision cosmology}",
      journal = {\physrep},
         year = 2015,
        month = feb,
       volume = {558},
        pages = {1-59},
          doi = {10.1016/j.physrep.2014.11.001},
archivePrefix = {arXiv},
       eprint = {1407.6990},
 primaryClass = {astro-ph.CO},
       adsurl = {https://ui.adsabs.harvard.edu/abs/2015PhR...558....1T}
}

@ARTICLE{2015SSRv..193...67K,
       author = {{Kiessling}, Alina and {Cacciato}, Marcello and {Joachimi}, Benjamin and {Kirk}, Donnacha and {Kitching}, Thomas D. and {Leonard}, Adrienne and {Mandelbaum}, Rachel and {Sch{\"a}fer}, Bj{\"o}rn Malte and {Sif{\'o}n}, Crist{\'o}bal and {Brown}, Michael L. and {Rassat}, Anais},
        title = "{Galaxy Alignments: Theory, Modelling \& Simulations}",
      journal = {\ssr},
         year = 2015,
        month = nov,
       volume = {193},
       number = {1-4},
        pages = {67-136},
          doi = {10.1007/s11214-015-0203-6},
archivePrefix = {arXiv},
       eprint = {1504.05546},
 primaryClass = {astro-ph.GA},
       adsurl = {https://ui.adsabs.harvard.edu/abs/2015SSRv..193...67K}
}

@ARTICLE{2020MNRAS.495.2436R,
       author = {{Ragone-Figueroa}, C. and {Granato}, G.~L. and {Borgani}, S. and {De Propris}, R. and {Garc{\'\i}a Lambas}, D. and {Murante}, G. and {Rasia}, E. and {West}, M.},
        title = "{Evolution and role of mergers in the BCG-cluster alignment. A view from cosmological hydrosimulations}",
      journal = {\mnras},
         year = 2020,
        month = jun,
       volume = {495},
       number = {2},
        pages = {2436-2445},
          doi = {10.1093/mnras/staa1389},
archivePrefix = {arXiv},
       eprint = {2004.13038},
 primaryClass = {astro-ph.GA},
       adsurl = {https://ui.adsabs.harvard.edu/abs/2020MNRAS.495.2436R}
}

@ARTICLE{2020MNRAS.495.3002K,
       author = {{Knebe}, Alexander and {G{\'a}mez-Mar{\'\i}n}, Mat{\'\i}as and {Pearce}, Frazer R. and {Cui}, Weiguang and {Hoffmann}, Kai and {De Petris}, Marco and {Power}, Chris and {Haggar}, Roan and {Mostoghiu}, Robert},
        title = "{The Three Hundred project: shapes and radial alignment of satellite, infalling, and backsplash galaxies}",
      journal = {\mnras},
         year = 2020,
        month = jul,
       volume = {495},
       number = {3},
        pages = {3002-3013},
          doi = {10.1093/mnras/staa1407},
archivePrefix = {arXiv},
       eprint = {2005.09896},
 primaryClass = {astro-ph.GA},
       adsurl = {https://ui.adsabs.harvard.edu/abs/2020MNRAS.495.3002K}
}

@ARTICLE{1970IrAJ....9..211O,
       author = {{{\"O}pik}, E.~J.},
        title = "{Preferential Orientation of Galaxies: on the Possibility of Detection}",
      journal = {Irish Astronomical Journal},
         year = 1970,
        month = mar,
       volume = {9},
        pages = {211},
       adsurl = {https://ui.adsabs.harvard.edu/abs/1970IrAJ....9..211O}
}

@ARTICLE{1968PASP...80..252S,
       author = {{Sastry}, Gummuluru N.},
        title = "{Clusters Associated with Supergiant Galaxies}",
      journal = {\pasp},
         year = 1968,
        month = jun,
       volume = {80},
       number = {474},
        pages = {252},
          doi = {10.1086/128626},
       adsurl = {https://ui.adsabs.harvard.edu/abs/1968PASP...80..252S}
}

@ARTICLE{1989ApJ...336...46W,
       author = {{West}, Michael J. and {Dekel}, Avishai and {Oemler}, Jr., Augustus},
        title = "{Cosmological Alignment of Clusters of Galaxies with Their Surroundings: A Problem for Cold Dark Matter?}",
      journal = {\apj},
         year = 1989,
        month = jan,
       volume = {336},
        pages = {46},
          doi = {10.1086/166991},
       adsurl = {https://ui.adsabs.harvard.edu/abs/1989ApJ...336...46W}
}

@ARTICLE{1993ApJ...418..544V,
       author = {{van Haarlem}, Michiel and {van de Weygaert}, Rien},
        title = "{Velocity Fields and Alignments of Clusters in Gravitational Instability Scenarios}",
      journal = {\apj},
         year = 1993,
        month = dec,
       volume = {418},
        pages = {544},
          doi = {10.1086/173416},
       adsurl = {https://ui.adsabs.harvard.edu/abs/1993ApJ...418..544V}
}

@ARTICLE{1997ApJ...479..632S,
       author = {{Splinter}, Randall J. and {Melott}, Adrian L. and {Linn}, Angela M. and {Buck}, Charles and {Tinker}, Jeremy},
        title = "{The Ellipticity and Orientation of Clusters of Galaxies in N-Body Experiments}",
      journal = {\apj},
         year = 1997,
        month = apr,
       volume = {479},
       number = {2},
        pages = {632-641},
          doi = {10.1086/303896},
archivePrefix = {arXiv},
       eprint = {astro-ph/9607144},
 primaryClass = {astro-ph},
       adsurl = {https://ui.adsabs.harvard.edu/abs/1997ApJ...479..632S}
}

@ARTICLE{2018MNRAS.481..414G,
       author = {{Ganeshaiah Veena}, Punyakoti and {Cautun}, Marius and {van de Weygaert}, Rien and {Tempel}, Elmo and {Jones}, Bernard J.~T. and {Rieder}, Steven and {Frenk}, Carlos S.},
        title = "{The Cosmic Ballet: spin and shape alignments of haloes in the cosmic web}",
      journal = {\mnras},
         year = 2018,
        month = nov,
       volume = {481},
       number = {1},
        pages = {414-438},
          doi = {10.1093/mnras/sty2270},
archivePrefix = {arXiv},
       eprint = {1805.00033},
 primaryClass = {astro-ph.CO},
       adsurl = {https://ui.adsabs.harvard.edu/abs/2018MNRAS.481..414G}
}

@ARTICLE{2018MNRAS.473.4077P,
       author = {{Pillepich}, Annalisa and {Springel}, Volker and {Nelson}, Dylan and {Genel}, Shy and {Naiman}, Jill and {Pakmor}, R{\"u}diger and {Hernquist}, Lars and {Torrey}, Paul and {Vogelsberger}, Mark and {Weinberger}, Rainer and {Marinacci}, Federico},
        title = "{Simulating galaxy formation with the IllustrisTNG model}",
      journal = {\mnras},
         year = 2018,
        month = jan,
       volume = {473},
       number = {3},
        pages = {4077-4106},
          doi = {10.1093/mnras/stx2656},
archivePrefix = {arXiv},
       eprint = {1703.02970},
 primaryClass = {astro-ph.GA},
       adsurl = {https://ui.adsabs.harvard.edu/abs/2018MNRAS.473.4077P}
}

@ARTICLE{2025ApJ...992..157G,
       author = {{Guo}, Yanhan and {Ma}, Qinglin and {Li}, Cheng},
        title = "{Lopsided and Bulging Distribution of Satellites Around Paired Halos. I. Observational Measurements and Comparison with Halo-based Models}",
      journal = {\apj},
         year = 2025,
        month = oct,
       volume = {992},
       number = {1},
          eid = {157},
        pages = {157},
          doi = {10.3847/1538-4357/ae00c3},
archivePrefix = {arXiv},
       eprint = {2502.14666},
 primaryClass = {astro-ph.GA},
       adsurl = {https://ui.adsabs.harvard.edu/abs/2025ApJ...992..157G}
}

@ARTICLE{2025ApJ...992...41M,
       author = {{Ma}, Qinglin and {Li}, Cheng and {Guo}, Yanhan},
        title = "{Lopsided and Bulging Distribution of Satellites around Paired Halos. II. 3D Analysis and Dependence on Projection and Selection Effects}",
      journal = {\apj},
         year = 2025,
        month = oct,
       volume = {992},
       number = {1},
          eid = {41},
        pages = {41},
          doi = {10.3847/1538-4357/adfede},
archivePrefix = {arXiv},
       eprint = {2502.14668},
 primaryClass = {astro-ph.CO},
       adsurl = {https://ui.adsabs.harvard.edu/abs/2025ApJ...992...41M}
}

@ARTICLE{2025MNRAS.544.3350L,
       author = {{Liu}, Shufei and {Zou}, Hu and {Gou}, Jinfu and {Guo}, Weijian and {Li}, Niu and {Li}, Wenxiong and {Singh}, Gaurav and {Song}, Haoming and {Sui}, Jipeng and {Tan}, Xi and {Xiao}, Yunao and {Zhang}, Jingyi and {Feng}, Lu},
        title = "{Galaxy clusters from the DESI legacy imaging surveys ─ III. Star-forming fraction of brightest cluster galaxies}",
      journal = {\mnras},
         year = 2025,
        month = dec,
       volume = {544},
       number = {4},
        pages = {3350-3360},
          doi = {10.1093/mnras/staf1885},
archivePrefix = {arXiv},
       eprint = {2511.09858},
 primaryClass = {astro-ph.GA},
       adsurl = {https://ui.adsabs.harvard.edu/abs/2025MNRAS.544.3350L}
}

@ARTICLE{2025ApJ...995..103C,
       author = {{Chen}, Zhaobin and {Gu}, Yizhou and {Zou}, Hu and {Yuan}, Qirong},
        title = "{Environmental Effects and Size─Mass Relations of Red Disk Galaxies at 0.1 < z < 0.5}",
      journal = {\apj},
         year = 2025,
        month = dec,
       volume = {995},
       number = {1},
          eid = {103},
        pages = {103},
          doi = {10.3847/1538-4357/ae18cf},
       adsurl = {https://ui.adsabs.harvard.edu/abs/2025ApJ...995..103C}
}

@ARTICLE{2025ApJ...990..171C,
       author = {{Chen}, Xinyue and {Xu}, Weiwei and {Li}, Ran and {Shan}, Huanyuan and {Yao}, Ji and {Wang}, Chunxiang},
        title = "{A Comparative Study of Halo Mass Estimates from Group Catalogs and Lensing Signals}",
      journal = {\apj},
         year = 2025,
        month = sep,
       volume = {990},
       number = {2},
          eid = {171},
        pages = {171},
          doi = {10.3847/1538-4357/adf332},
archivePrefix = {arXiv},
       eprint = {2507.20294},
 primaryClass = {astro-ph.CO},
       adsurl = {https://ui.adsabs.harvard.edu/abs/2025ApJ...990..171C}
}

@ARTICLE{2011Natur.477..567W,
       author = {{Wojtak}, Rados{\l}aw and {Hansen}, Steen H. and {Hjorth}, Jens},
        title = "{Gravitational redshift of galaxies in clusters as predicted by general relativity}",
      journal = {\nat},
         year = 2011,
        month = sep,
       volume = {477},
       number = {7366},
        pages = {567-569},
          doi = {10.1038/nature10445},
archivePrefix = {arXiv},
       eprint = {1109.6571},
 primaryClass = {astro-ph.CO},
       adsurl = {https://ui.adsabs.harvard.edu/abs/2011Natur.477..567W}
}

@ARTICLE{2012MNRAS.424..232W,
       author = {{Wetzel}, Andrew R. and {Tinker}, Jeremy L. and {Conroy}, Charlie},
        title = "{Galaxy evolution in groups and clusters: star formation rates, red sequence fractions and the persistent bimodality}",
      journal = {\mnras},
         year = 2012,
        month = jul,
       volume = {424},
       number = {1},
        pages = {232-243},
          doi = {10.1111/j.1365-2966.2012.21188.x},
archivePrefix = {arXiv},
       eprint = {1107.5311},
 primaryClass = {astro-ph.CO},
       adsurl = {https://ui.adsabs.harvard.edu/abs/2012MNRAS.424..232W}
}

@ARTICLE{2007MNRAS.378.1531K,
       author = {{Kang}, X. and {van den Bosch}, Frank C. and {Yang}, Xiaohu and {Mao}, Shude and {Mo}, H.~J. and {Li}, Cheng and {Jing}, Y.~P.},
        title = "{The alignment between satellites and central galaxies: theory versus observations}",
      journal = {\mnras},
         year = 2007,
        month = jul,
       volume = {378},
       number = {4},
        pages = {1531-1542},
          doi = {10.1111/j.1365-2966.2007.11902.x},
archivePrefix = {arXiv},
       eprint = {astro-ph/0701130},
 primaryClass = {astro-ph},
       adsurl = {https://ui.adsabs.harvard.edu/abs/2007MNRAS.378.1531K}
}

@ARTICLE{2021Natur.594..187M,
       author = {{Mart{\'\i}n-Navarro}, Ignacio and {Pillepich}, Annalisa and {Nelson}, Dylan and {Rodriguez-Gomez}, Vicente and {Donnari}, Martina and {Hernquist}, Lars and {Springel}, Volker},
        title = "{Anisotropic satellite galaxy quenching modulated by black hole activity}",
      journal = {\nat},
         year = 2021,
        month = jun,
       volume = {594},
       number = {7862},
        pages = {187-190},
          doi = {10.1038/s41586-021-03545-9},
archivePrefix = {arXiv},
       eprint = {2106.04587},
 primaryClass = {astro-ph.GA},
       adsurl = {https://ui.adsabs.harvard.edu/abs/2021Natur.594..187M}
}

@ARTICLE{2009Natur.460..213C,
       author = {{Cattaneo}, A. and {Faber}, S.~M. and {Binney}, J. and {Dekel}, A. and {Kormendy}, J. and {Mushotzky}, R. and {Babul}, A. and {Best}, P.~N. and {Br{\"u}ggen}, M. and {Fabian}, A.~C. and {Frenk}, C.~S. and {Khalatyan}, A. and {Netzer}, H. and {Mahdavi}, A. and {Silk}, J. and {Steinmetz}, M. and {Wisotzki}, L.},
        title = "{The role of black holes in galaxy formation and evolution}",
      journal = {\nat},
         year = 2009,
        month = jul,
       volume = {460},
       number = {7252},
        pages = {213-219},
          doi = {10.1038/nature08135},
archivePrefix = {arXiv},
       eprint = {0907.1608},
 primaryClass = {astro-ph.CO},
       adsurl = {https://ui.adsabs.harvard.edu/abs/2009Natur.460..213C}
}

@ARTICLE{2025ApJ...992L..17W,
       author = {{Wang}, Peng},
        title = "{Observational Evidence for Spin Alignment between Galaxy Groups and Their Central Galaxies}",
      journal = {\apjl},
         year = 2025,
        month = oct,
       volume = {992},
       number = {1},
          eid = {L17},
        pages = {L17},
          doi = {10.3847/2041-8213/ae0ca5},
archivePrefix = {arXiv},
       eprint = {2509.24278},
 primaryClass = {astro-ph.GA},
       adsurl = {https://ui.adsabs.harvard.edu/abs/2025ApJ...992L..17W}
}

@ARTICLE{2025ApJ...987L..30W,
       author = {{Wang}, Hao-da and {Wang}, Peng and {Bao}, Min and {Chen}, Yanmei and {Tang}, Xiao-xiao and {Zhang}, Youcai and {Kang}, Xi and {Guo}, Quan and {Sheng}, Ming-Jie and {Yu}, Hao-Ran},
        title = "{Observed Antiparallel Correlation between Spiral Galaxy and Cosmic Filament Spins}",
      journal = {\apjl},
         year = 2025,
        month = jul,
       volume = {987},
       number = {2},
          eid = {L30},
        pages = {L30},
          doi = {10.3847/2041-8213/ade6fe},
archivePrefix = {arXiv},
       eprint = {2506.22794},
 primaryClass = {astro-ph.GA},
       adsurl = {https://ui.adsabs.harvard.edu/abs/2025ApJ...987L..30W}
}

@ARTICLE{2008ApJ...672..825P,
       author = {{Pereira}, Maria J. and {Bryan}, Greg L. and {Gill}, Stuart P.~D.},
        title = "{Radial Alignment in Simulated Clusters}",
      journal = {\apj},
         year = 2008,
        month = jan,
       volume = {672},
       number = {2},
        pages = {825-833},
          doi = {10.1086/523830},
archivePrefix = {arXiv},
       eprint = {0707.1702},
 primaryClass = {astro-ph},
       adsurl = {https://ui.adsabs.harvard.edu/abs/2008ApJ...672..825P}
}

@ARTICLE{2020ApJ...900..129W,
       author = {{Wang}, Peng and {Libeskind}, Noam I. and {Tempel}, Elmo and {Pawlowski}, Marcel S. and {Kang}, Xi and {Guo}, Quan},
        title = "{The Alignment of Satellite Systems with Cosmic Filaments in the SDSS DR12}",
      journal = {\apj},
         year = 2020,
        month = sep,
       volume = {900},
       number = {2},
          eid = {129},
        pages = {129},
          doi = {10.3847/1538-4357/aba6ea},
archivePrefix = {arXiv},
       eprint = {2007.08345},
 primaryClass = {astro-ph.GA},
       adsurl = {https://ui.adsabs.harvard.edu/abs/2020ApJ...900..129W}
}

@ARTICLE{2024MNRAS.529.3651O,
       author = {{Oxland}, M. and {Parker}, L.~C. and {de Carvalho}, R.~R. and {Sampaio}, V.~M.},
        title = "{Satellite quenching and morphological transformation of galaxies in groups and clusters}",
      journal = {\mnras},
         year = 2024,
        month = apr,
       volume = {529},
       number = {4},
        pages = {3651-3665},
          doi = {10.1093/mnras/stae747},
archivePrefix = {arXiv},
       eprint = {2403.07742},
 primaryClass = {astro-ph.GA},
       adsurl = {https://ui.adsabs.harvard.edu/abs/2024MNRAS.529.3651O}
}

@ARTICLE{2023MNRAS.518.2398H,
       author = {{Hough}, Tom{\'a}s and {Cora}, Sof{\'\i}a A. and {Haggar}, Roan and {Vega-Martinez}, Cristian and {Kuchner}, Ulrike and {Pearce}, Frazer and {Gray}, Meghan and {Knebe}, Alexander and {Yepes}, Gustavo},
        title = "{The Three Hundred Project: Connection between star formation quenching and dynamical evolution in and around simulated galaxy clusters}",
      journal = {\mnras},
         year = 2023,
        month = jan,
       volume = {518},
       number = {2},
        pages = {2398-2417},
          doi = {10.1093/mnras/stac3209},
archivePrefix = {arXiv},
       eprint = {2211.04485},
 primaryClass = {astro-ph.GA},
       adsurl = {https://ui.adsabs.harvard.edu/abs/2023MNRAS.518.2398H}
}

@ARTICLE{2010ApJ...721..939P,
       author = {{Pereira}, Maria J. and {Bryan}, Greg L.},
        title = "{Tidal Torquing of Elliptical Galaxies in Cluster Environments}",
      journal = {\apj},
         year = 2010,
        month = oct,
       volume = {721},
       number = {2},
        pages = {939-955},
          doi = {10.1088/0004-637X/721/2/939},
archivePrefix = {arXiv},
       eprint = {1009.4191},
 primaryClass = {astro-ph.CO},
       adsurl = {https://ui.adsabs.harvard.edu/abs/2010ApJ...721..939P}
}

@ARTICLE{2026AJ....171..285D,
       author = {{DESI Collaboration} and {Abdul Karim}, M. and {Adame}, A.~G. and {Aguado}, D. and {Aguilar}, J. and {Ahlen}, S. and {Alam}, S. and {Aldering}, G. and {Alexander}, D.~M. and {Alfarsy}, R. and et al.},
        title = "{Data Release 1 of the Dark Energy Spectroscopic Instrument}",
      journal = {\aj},
         year = 2026,
        month = may,
       volume = {171},
       number = {5},
          eid = {285},
        pages = {285},
          doi = {10.3847/1538-3881/ae4c43},
archivePrefix = {arXiv},
       eprint = {2503.14745},
 primaryClass = {astro-ph.CO},
       adsurl = {https://ui.adsabs.harvard.edu/abs/2026AJ....171..285D}
}

@ARTICLE{2008MNRAS.388.1321P,
       author = {{Padilla}, Nelson D. and {Strauss}, Michael A.},
        title = "{The shapes of galaxies in the Sloan Digital Sky Survey}",
      journal = {\mnras},
         year = 2008,
        month = aug,
       volume = {388},
       number = {3},
        pages = {1321-1334},
          doi = {10.1111/j.1365-2966.2008.13480.x},
archivePrefix = {arXiv},
       eprint = {0802.0877},
 primaryClass = {astro-ph},
       adsurl = {https://ui.adsabs.harvard.edu/abs/2008MNRAS.388.1321P}
}

@ARTICLE{2018ApJ...859..115W,
       author = {{Wang}, Peng and {Luo}, Yu and {Kang}, Xi and {Libeskind}, Noam I. and {Wang}, Lei and {Zhang}, Youcai and {Tempel}, Elmo and {Guo}, Quan},
        title = "{Alignment between Satellite and Central Galaxies in the SDSS DR7: Dependence on Large-scale Environment}",
      journal = {\apj},
         year = 2018,
        month = jun,
       volume = {859},
       number = {2},
          eid = {115},
        pages = {115},
          doi = {10.3847/1538-4357/aabe2b},
archivePrefix = {arXiv},
       eprint = {1802.10105},
 primaryClass = {astro-ph.GA},
       adsurl = {https://ui.adsabs.harvard.edu/abs/2018ApJ...859..115W}
}

@ARTICLE{2010ApJ...725..794Q,
       author = {{Quadri}, Ryan F. and {Williams}, Rik J.},
        title = "{Quantifying Photometric Redshift Errors in the Absence of Spectroscopic Redshifts}",
      journal = {\apj},
         year = 2010,
        month = dec,
       volume = {725},
       number = {1},
        pages = {794-802},
          doi = {10.1088/0004-637X/725/1/794},
archivePrefix = {arXiv},
       eprint = {0910.2704},
 primaryClass = {astro-ph.CO},
       adsurl = {https://ui.adsabs.harvard.edu/abs/2010ApJ...725..794Q}
}
\bibliographystyle{aasjournalv7}

\end{CJK*}
\end{document}